\documentclass[fleqn,usenatbib]{mnras}

\usepackage{newtxtext,newtxmath}

\usepackage[T1]{fontenc}

\DeclareRobustCommand{\VAN}[3]{#2}
\let\VANthebibliography\thebibliography
\def\thebibliography{\DeclareRobustCommand{\VAN}[3]{##3}\VANthebibliography}

\usepackage{graphicx}	
\usepackage{amsmath}	
\usepackage{relsize}

\graphicspath{ {./images/} }
\usepackage{bm}
\usepackage{multirow} 
\usepackage{gensymb} 
\usepackage{placeins} 

\defcitealias{EHT1}{EHT~I}
\defcitealias{EHT2}{EHT~II}
\defcitealias{EHT3}{EHT~III}
\defcitealias{EHT4}{EHT~IV}
\defcitealias{EHT5}{EHT~V}
\defcitealias{EHT6}{EHT~VI}

\title[M87* Likelihood]{
How narrow is the M87* ring?
\\
I. The choice of closure likelihood function
}

\author[
Lockhart and Gralla
]{
Will Lockhart$^{1}$\thanks{wlockhart@email.arizona.edu}
and Samuel E. Gralla,$^{1}$
\\
$^{1}$Department of Physics, University of Arizona, Tucson AZ, USA
}

\begin{document}
\label{firstpage}
\pagerange{\pageref{firstpage}--\pageref{lastpage}}
\maketitle

\begin{abstract}

Event Horizon Telescope (EHT) observations of the core of the galaxy M87 suggest an observational appearance dominated by a ring of approximately 40$\mu$as in diameter.  The thickness of the ring is less certain: imaging efforts constrained it to be less than half the diameter (consistent with an imaging resolution of 20$\mu$as), while visibility-domain modeling suggested a variety of fractional widths, including as low as $10\%$ on some days.  The fractional width is very interesting as it has the potential to discriminate between different astrophysical scenarios for the source; in fact, the $10$--$20\%$ range is so narrow as to be in tension with theoretical expectations.  In the first of a series of papers on the width of the observed ring, we reproduce a subset of EHT visibility-domain modeling results and we explore whether alternative data analysis methods might favor thicker rings.  We point out that the closure phase (and closure amplitude) likelihood function is not independent of residual station gain amplitudes, even at high signal-to-noise, and explore two approximations of practical interest: one standard in the field (and employed by the EHT collaboration), and a new one that we propose.  Analyzing the public data, we find that the new likelihood approximation prefers somewhat thicker rings, more in line with theoretical expectations.  Further analysis is needed, however, to determine which approximation is better for the EHT data.
\vspace{1mm}
\end{abstract}

\begin{keywords}
galaxies: nuclei -- (galaxies:) quasars: supermassive black holes -- submillimetre: galaxies -- black hole physics -- techniques: interferometric
\vspace{-3mm}
\end{keywords}

\section{Introduction}

The 2017 Event Horizon Telescope (EHT) observations are the first electromagnetic observations probing event horizon scales of a black hole, a remarkable achievement that will undoubtedly be remembered as a milestone in observational astronomy \citep[henceforth EHT I-VI]{EHT1,EHT2,EHT3,EHT4,EHT5,EHT6}.  However, as recognized by the collaboration, there are many challenges in interpreting the data, beginning with the most basic question of the observational appearance of the source.  These difficulties arise because the data do not directly probe the sky brightness, but rather provide a sparse sampling of its Fourier transform (the ``complex visibility'') subject to a high degree of calibration uncertainty.  To infer the observational appearance, the EHT collaboration (EHTC) tried a variety of different approaches and found that all supported the presence of a ring of approximately $40$ micro-arcseconds ($\mu\text{as}$) in diameter, with a mild brightness gradient roughly in the southerly direction.  This matches expectations for matter orbiting near the event horizon of a black hole weighing several billion solar masses \citep{m87-mass-stellar,m87-mass-gas}, supporting the prevailing model for this type of source \citep{melia1992,narayan-yi1994,falcke-melia-agol2000} and clinching the black hole interpretation of M87*.

The identification of a robust feature matching expectations is a great success, but many interesting questions remain.  What is the precise orientation (position angle) of the brightness gradient?  How wide is the ring?  How quickly does the brightness fall off with distance from the black hole?  What other image features are present?  Until such questions can be answered, efforts to infer more detailed source properties (i.e., to ``do astrophysics'' with these observations) will be severely hampered.  While EHTC certainly \textit{asked} many of these questions in their analysis, the results were unfortunately inconclusive, in that the answers varied among techniques and datasets.  While it may be that more data is ultimately needed for conclusive answers, we hope to contribute to the conversation by providing our own interpretation of this remarkable first dataset. 

\begin{figure}
    \centering
    \includegraphics[width=\linewidth]{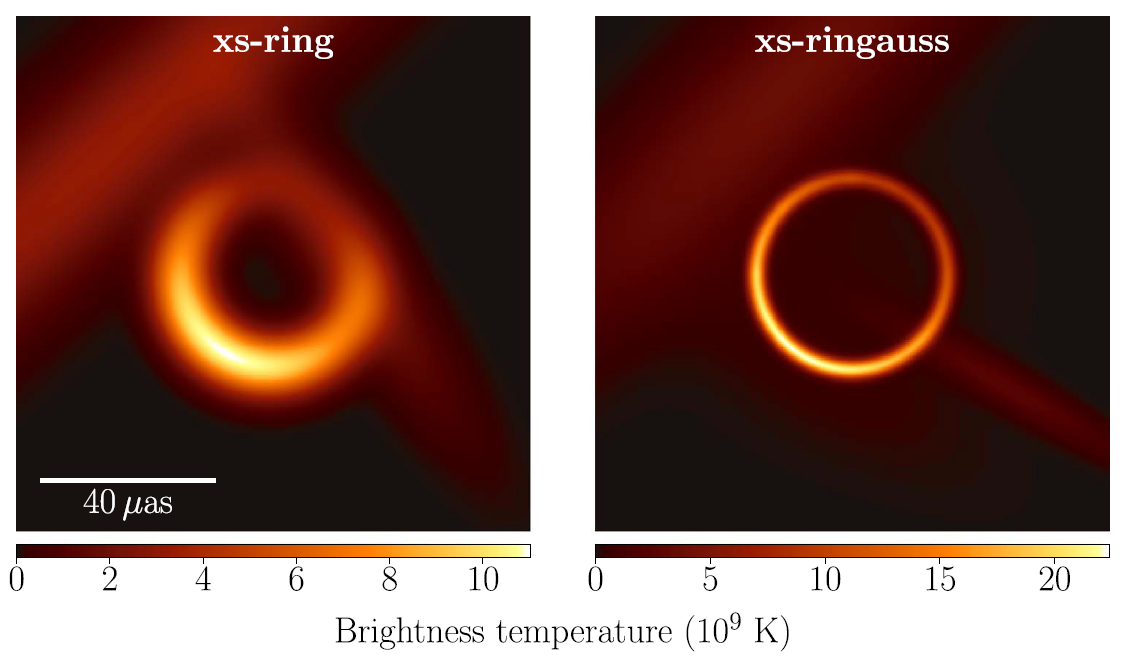}
    \caption{Typical sky appearance favored by EHTC visibility-domain modeling for the Apr 6 hi-band dataset (reproduced from Fig.~5 of \citetalias{EHT6}).  Such narrow rings are in tension with theoretical expectations (see discussion in Sec.~\ref{sec:theory} and App.~\ref{app:mechanisms}).}
    \label{fig:rings}
\end{figure}

We are motivated in particular by the question of the \textit{thickness} of the ring.  EHTC reported only an upper bound on the ring width, finding that it must be less than half the diameter (fractional width $\lesssim50\%$).  This lack of constraining power is unfortunate because the ring width is a useful probe of astrophysical source models, providing direct information about the emission profile.  Furthermore, some of the most detailed modeling highlighted by EHTC suggested extremely narrow rings, with fractional widths of only $10-20\%$ (Fig.~\ref{fig:rings}).  Rings this narrow would seem to require  rather dramatic revisions in ideas about the origin of the 1.3mm emission from M87* (App.~\ref{app:mechanisms}).\footnote{The gravitationally lensed ``photon ring'', while very thin, is always subdominant to broader direct emission, and therefore cannot explain a narrow ring overall.  \citep{gralla-holz-wald2019,johnson-etal2020,gralla2021,chael-johnson-lupsasca2021,broderick-tiede-pesce-gold2021}.}

In our view, trying to pin down the width of the ring feature is the natural next step after the definitive results on its diameter.  In the first of a series of papers on this issue, we explore whether improvements in data analysis might modify the favored fractional widths.  In particular, we identify an implicit (and apparently standard) approximation made by EHTC in computing likelihood functions for the closure phases and closure amplitudes that are used to mitigate calibration errors.  The issue is that the while calibration uncertainties (``residual station gains'') cancel out from the \textit{mean values} of the closure phases and closure amplitudes predicted by a given model, they remain present in the \textit{variances}.  These variances contribute to the likelihood, and some approximation must be made if the likelihood is to be independent of station gains \citep{blackburn2020}.

The EHTC approximation consists of replacing the gain-modified model mean value with the measured mean value when calculating the variances.  We propose an alternative where one simply drops the residual station gain amplitudes, using the \textit{model} mean value where EHTC had used the \textit{measured} mean value.  In the EHTC approximation the variances may be computed once and for all from the data, and we refer to this as the ``fixed'' likelihood.  In the new approximation the variances depend on the model parameters, and we refer to this as the ``variable'' likelihood.  The fixed likelihood (EHTC) is expected to perform better in the limit of large residual amplitude gains, while the variable likelihood (this paper) is expected to perform better in the limit of small amplitude gains (see Sec.~\ref{sec:closure}).  We see no reason to prefer one over the other for the EHT dataset.  

\begin{figure*}
    \centering
    \includegraphics[width=0.95\textwidth]{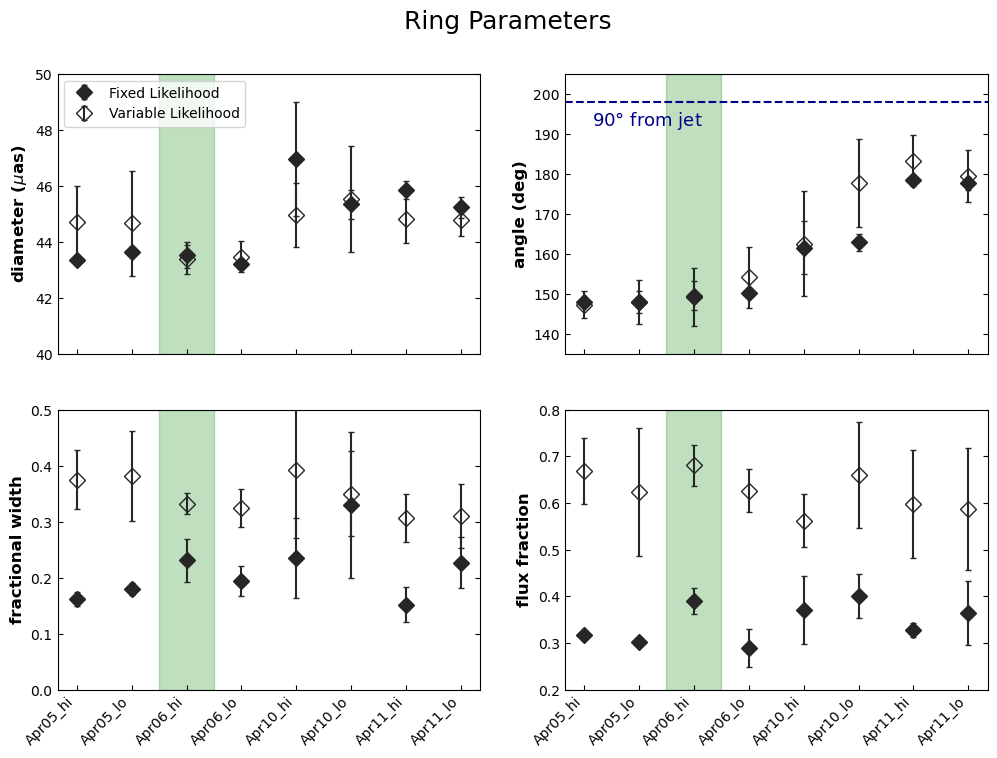}
    \caption{Summary of the main results.  In our analysis of the 8 datasets released by EHT (April 05, 06, 10, 11 in low and high bands), we focus on four parameters characterizing the ring: the diameter, the fractional width (width divided by diameter), the position angle of the brighest part, and the fraction of the total flux (ring flux divided by total flux, where total flux includes nuissance Gaussian components).  Our analysis makes a number of arbitrary choices involving the number of nuissance parameters, the sampling method used, and the handling of low-SNR data points.  On these plots, a point with an error bar indicates the mean and standard deviation \textit{of the mean values} found in the  different runs with different arbitrary choices.  (See Fig.~\ref{fig:apr06hi_results} for the individual posterior ranges for the dataset highlighted here in green.)  The diameter is consistent among methods and days.  The angle shows some variation towards the angle corresponding to matter orbiting in the plane perpendicular to the jet with angular momentum proportional to the counter-jet (blue line).  The fractional width and flux fraction depend on the choice of fixed versus variable likelihood approximation.  This highlights the need to better understand the reliability of these approximations.}
    \label{fig:master_plot}
\end{figure*}
  
As a first step in quantifying the importance of the approximation, we will reconsider one subset the EHTC analysis: a geometric image model known as \textsf{xs-ring}, fit using closure phases and closure amplitudes.  Working from the EHT public data, but with a new analysis pipeline built from scratch, we reproduce the collaboration's fixed-likelihood results and provide analogous results for the variable likelihood (Fig.~\ref{fig:master_plot}).  We find that the fixed likelihood prefers fractional widths of around $20\%$, while the variable likelihood prefers fractional widths of around $35\%$.  We conclude that the choice of likelihood approximation \textit{does affect the results}, and the new choice gives results more consistent with theoretical expectations.  However, we emphasize that further analysis is required to determine which likelihood approximation is more reliable for the EHT dataset.  

In addition to the fractional width, we pay particular attention to the ring diameter, position angle (orientation of brightness gradient), and the fraction of the flux in the ring component (Fig.~\ref{fig:master_plot}).  (The image also contains flux from additional ``nuisance Gaussian'' components added to the \textsf{xs-ring} model).  We find that the diameter and position angle are roughly independent of the choice of likelihood (fixed or variable), and broadly consistent with EHTC's reported values for \textsf{xs-ring}.  The position angle shows a clear trend over the several days of observation, increasing towards (but not quite reaching) the value corresponding to relativistic matter orbiting in the plane perpendicular to the M87 jet.\footnote{The jet is oriented at $288 \degree$ projected along the line of sight and angled $17 \degree$ relative to the line of sight.  If matter orbits in the plane perpendicular to the jet, the brightest part of the observed emission will be at $288\pm 90 \degree$, depending on the sense of rotation.  The value $198 \degree$ corresponds to orbits in the left-handed sense with the thumb pointed toward the jet, i.e., angular momentum proportional to the counter-jet.\label{foot:jet}} By contrast, we find that the flux fraction changes significantly with the choice of likelihood approximation.  This further highlights the need for exploring both formulations.

Beyond reporting these results, we have three goals in mind for this paper.  First, by reporting our independent verification of a subset of the EHTC data analysis, we hope to increase confidence in the EHT results.  Although there was no reason to expect an error in the EHTC analysis, we feel that independent verification is essential to the scientific method, especially in an environment where a single, large experiment operates without competition.  Second, by highlighting the utility of ring width as a probe of source properties, we hope to stimulate additional interest in a problem that could in principle result in a real revision in our understanding of M87*.  Third, by discussing subtleties in the choice of likelihood function, we hope to encourage future statistical analyses, either exploring these approximations or going beyond them.

This paper is organized as follows.  In Sec.~\ref{sec:narrowness} we review observational inference and theoretical expectations for M87*.  In Sec.~\ref{sec:closure} we discuss the fixed and variable approximations for the likelihood function in detail.  In Sec.~\ref{sec:data} we review the public EHT data and describe our analysis approach.  In Sec.~\ref{sec:analysis} we present our modeling results, and in Sec.~\ref{sec:discussion} we summarize our conclusions and outlook.  Several issues are expounded in the Appendices, which discuss (\ref{app:xs-ring}) the details of the \textsf{xs-ring} model; (\ref{app:like}) the construction of optimal closure sets; (\ref{app:goodness}) the notion of goodness of fit; (\ref{app:just_add_one}) an approximation for the impact parameter of emitted photons; and (\ref{app:mechanisms}) the relationship between source properties and observed ring width.


\section{Ring Width: Theory and Observation}\label{sec:narrowness}

\subsection{Observational Inference}\label{sec:observe}

The EHT data provide a sparse, imperfectly-calibrated sampling of the Fourier transform of the sky brightness.  As such, inference of the underlying image features is necessarily probabilistic, and in practice requires many arbitrary choices in order to make progress.  The approach of the EHT collaboration was to consider a number of different strategies and look for points of agreement among results.  Broadly, EHTC considered two classes of method: image reconstruction and geometric modeling.\footnote{EHTC also attempted to directly constrain a class of source models based on general-relativistic magnetohydrodynamics (GRMHD) simulations.  Here we focus on the question of the observational appearance, independent of an assumed source model.}  These are distinguished in that an imaging algorithm attempts to be agnostic to the underlying image appearance and provides a single best-guess image according to the specific algorithm, while the geometric modeling approach assumes a model for the underlying image and provides posterior probability distributions for its parameters.

EHTC considered three different imaging algorithms and two geometric models, exploring a large parameter freedom within these approaches.  All approaches produced an annulus with a typical diameter of 40$\mu$as,
\begin{align}
    d \approx 40\mu\textrm{as}.
\end{align}
However, the thickness of the annulus varied significantly with the technique.  In the April 6 hi-band dataset highlighted by EHT (see Fig.~16 of \citetalias{EHT6}), the imaging algorithms preferred thicker rings (fractional widths of 30--50\%), while geometric modeling preferred narrow rings of only 10--20\% the annulus diameter (Fig.~\ref{fig:rings}).  Although the difference is less severe on other days, the general preference for thinner rings in the geometric modeling results is generic across datasets.\footnote{By fractional width, we mean the angle-averaged full-width-half-maximum (FWHM) of the ring.  This definition is used in Fig.~16 of \citetalias{EHT6}.  The fractional width parameter $\hat{f}_w$ reported elsewhere in \citetalias{EHT6} is biased upwards from this value (Fig.~\ref{fig:FWHM} below).}  EHTC suggests that the geometric modeling results may be more reliable for fractional width estimates, as the widths measured from reconstructed images are  ``systematically biased upward from the true values,'' such that they can ``at best be viewed as upper limits'' \citepalias[Sec.~9]{EHT4}.  To summarize, EHTC analysis suggested widths in the range
\begin{align}
    5 \mu\textrm{as} \lesssim w \lesssim 20 \mu\textrm{as} \qquad (10\% \lesssim w/d \lesssim 50\%).
\end{align}

Although EHTC is clear on the stated limit $w/d < 0.5$ produced by their analysis, it is important to note that composite images will in general violate this bound.  For example, the images highlighted in \citetalias{EHT1} are produced by averaging results from different imaging methods after blurring each with $20\mu$as (or similar) Gaussian kernels.  The imaging algorithms themselves also involve Gaussian blurring, so that the images in \citetalias{EHT1} are, in effect, triply blurred. 


\subsection{Theoretical Interpretation}\label{sec:theory}

Under the assumption that the 1.3mm radiation observed by EHT is produced by matter surrounding a black hole, we may associate an emission region with the observed ring by following null geodesics in the Kerr spacetime.  The typical angular scale associated with a mass $M$ at a distance $D$ is
\begin{align}\label{thetaM}
   \alpha_M = \frac{G M}{D c^2} = \psi \cdot (3.62 \  \mu\textrm{as} ),
\end{align}
where the scaling factor $\psi$ is equal to unity for the canonical values $M=6.2\times 10^9 M_\odot$ and $D=16.9$ Mpc for the M87* mass and distance.  With the canonical distance, $\psi=1$ is the mass favored by stellar dynamical measurements \citep{m87-mass-stellar}, while $\psi=0.56$ is the mass favored by gas-dynamical measurements \citep{m87-mass-gas}.

\begin{figure}
    \centering
    \includegraphics[width=\linewidth]{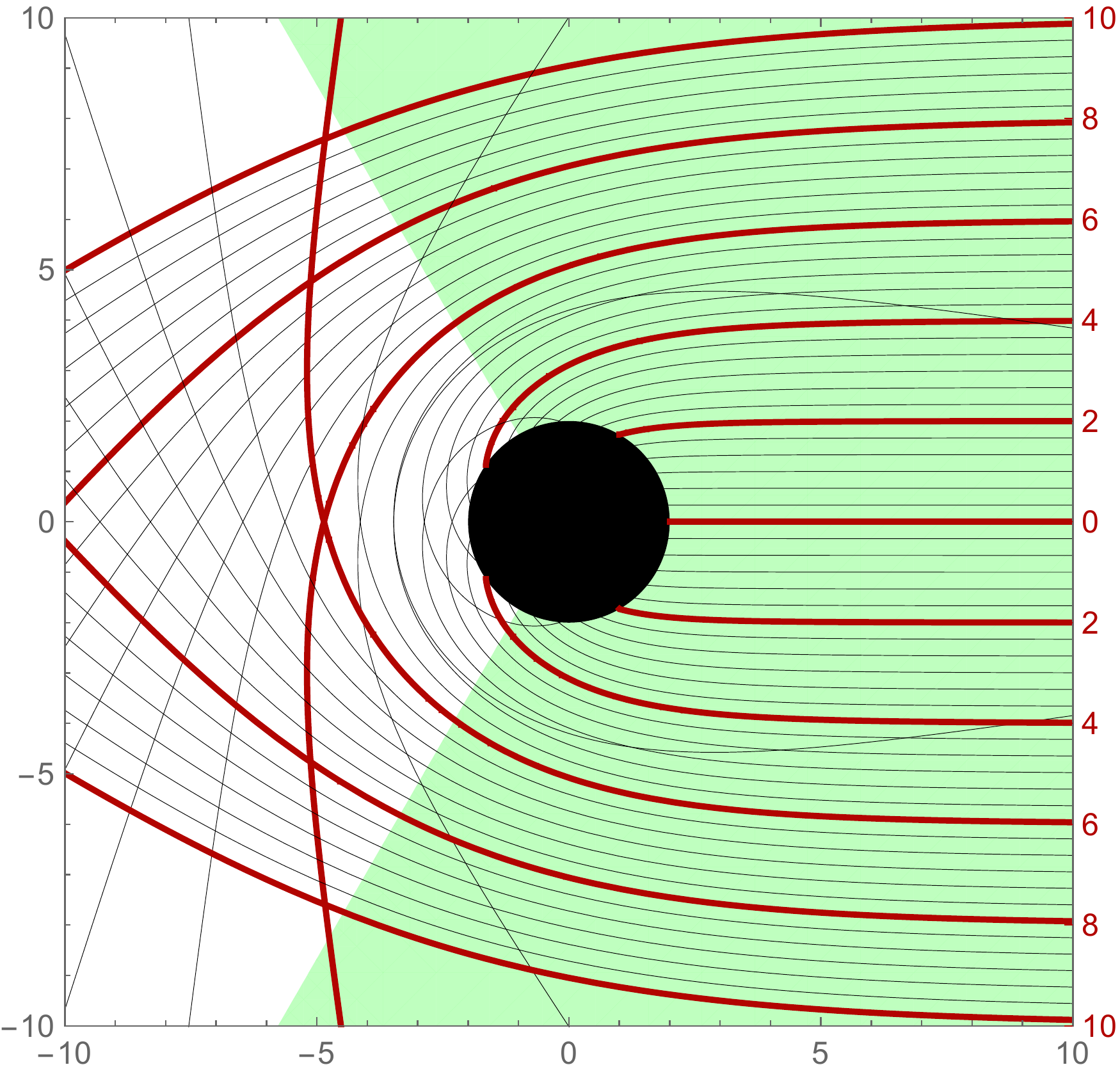}
    \caption{Bending of light by a Schwarzschild black hole, showing photon trajectories relevant to a distant observer on the right.  Quantities are shown in units of $GM/c^2$, with the impact parameter of selected rays (red) indicated on the right.  This dimensionless impact parameter may be multiplied by $\alpha_M$ \eqref{thetaM} to determine the associated angular radius on an image.   Sources in the green region are relatively undistorted by lensing, as quantified by the validity of a simple, ``just add one'' approximation (App.~\ref{app:just_add_one}) for all spin magnitude and orientation. This means that the observed ring width is a faithful probe of the typical dimensions of the emission profile.}
    \label{fig:conchy}
\end{figure}

Determining the observational appearance of a given emission profile requires solving the equations of radiative transport in the Kerr spacetime.  In the optically thick limit, each pixel on the image corresponds to a single emission point in the Kerr spacetime, whereas in the optically thin limit, each pixel reflects the integrated emission along a whole null geodesic.  A selection of the null geodesics relevant to a given observer is shown in Fig.~\ref{fig:conchy}.  This kind of plot can be used to build intuition for the observational appearance of a given source by sketching the source on the plot and following rays that emerge from it to their associated image radius $\alpha/\alpha_M$ shown in red on the right.  If the rays remain mostly parallel (irrespective of whether they bend), then the source is not significantly distorted by lensing, appearing on the image with the same relative dimensions it has in the source.

It is visually evident that the distortion is mostly negligible in the entire (green-colored) foregound/midplane region facing the observer.  This claim is quantified in appendix \ref{app:just_add_one} via the validity of a simple ``just add one'' formula for the arrival impact parameter of rays in this region, irrespective of spin and observer inclination.  The distortion is more significant in the white rear region, reaching a peak directly behind the black hole, where a line of emission extending from $2$ to $10$ is lensed into a ring ranging from $4$ to $8$, i.e., the typical dimensions are cut in half.  Rays that bend more than $90\degree$ carry secondary (further demagnified) images of the source; these arrive near a ``critical curve'' on the image plane \citep{bardeen1973} ($\alpha/\alpha_M\sim5$), creating a ``photon ring'' feature \citep{luminet1979} that does not affect the gross features of the image \citep{gralla-holz-wald2019}.

The key takeaway of this discussion is that \textit{apart from finely tuned special cases, gravitational lensing does not significantly modify the typical dimensions of the source}.  While emission from very near the black hole is always suppressed due to redshift-dimming, the width of the observed ring is otherwise a direct probe of the width of the emission profile.  This makes the ring width a promising observable for distinguishing between source models.  We discuss more details of this mapping between emission profile and observed image in App.~\ref{app:mechanisms}.

\section{Closure Likelihood Approximations}\label{sec:closure}
In this section, we review the basics of radio interferometry, introduce the closure quantities used to mitigate calibration uncertainty, and discuss subtleties in the choice of closure likelihood function.  Let $I(x,y)$ be the sky brightness at some particular wavelength (i.e., the specific intensity) as a function of (small) dimensionless direction cosines $x$ and $y$.  We will refer to $I(x,y)$ as ``the image''.  Its Fourier transform will be denoted $\mathcal{V}$,
\begin{align}\label{fourier_transform}
\mathcal{V}(u,v) = \int \!\!\! \int \textrm{e}^{-2 \pi i (ux + vy)} I(x,y) \, \textrm{d}x \, \textrm{d}y.
\end{align}
The fundamental observable of radio interferometry is the cross-correlation of the complex electric field at different stations $i$ and $j$, otherwise known as the complex visibility $\mathcal{V}_{ij}$.  Assuming that the source is slowly varying and spatially incoherent (properties implicit in its representation as an intensity $I(x,y)$), then the complex visibility samples the Fourier transform as
\begin{align}\label{vis}
    \mathcal{V}_{ij} = \mathcal{V}(u_{ij},v_{ij}), 
\end{align}
where the so-called ``baseline'' $(u_{ij},v_{ij})$ is a vector equal to the line-of-sight projected separation between stations $i$ and $j$, divided by the observational wavelength.  These assumptions are extremely well-satisfied for astronomical sources, and we will regard the association \eqref{vis} as exact.

The measurement of the complex electric field at each site is subject to thermal noise, which is Gaussian and easily estimated, together with calibration errors, which are much harder to control.  The resulting error budget for the complex visibility is \citepalias[Eq.~(2)]{EHT6}
\begin{align}\label{visbudget}
V_{ij} = g_i(t) g^*_j(t) \mathcal{V}_{ij} + \epsilon_{\sigma_{ij}},
\end{align}
where $g_i(t)$ are complex residual gains (unknown calibration factors) for each station $i$ as a function of time $t$, $\epsilon_{\sigma}$ is a mean-zero complex Gaussian random variable with standard deviation $\sigma$, and $\sigma_{ij}$ is the measured thermal noise.  We regard $V_{ij}$ as a random variable, one realization of which is reported in any given observation.  The specific measured value in some observation will be denoted with a hat, i.e.,
\begin{align}
    \hat{V}_{ij} = \textrm{(measured value of $V_{ij}$)}.
\end{align}

The presence of the residual station gains $g_i$ complicates the process of inferring the true visibility $\mathcal{V}_{ij}$ from the measured visibility $\hat{V}_{ij}$ and estimated noise $\sigma_{ij}$ provided by the EHT observations.  One approach is to simply include these coefficients as additional unknown parameters when fitting a model for $\mathcal{V}$. However, this introduces a large number of new parameters, since the gains generally vary on an atmospheric timescale, even if instrumental effects are negligible.  Collecting more data actually makes this problem worse, as the number of gain parameters scales linearly with both array size and observation time. 

An alternative approach is to focus on so-called ``closure quantities'' that are less sensitive to station gains \citep{TMS}.  A closure phase is constructed from three stations $\{i,j,k\}$ by 
\begin{align}\label{closure_phase}
    \psi_{ijk} = \arg (V_{ij}V_{jk}V_{ki}) = \phi_{ij} +\phi_{jk} +\phi_{ki},
\end{align}
where $\phi_{ij} = \arg(V_{ij})$ is the phase of each visibility (with phase arithmetic mod $2\pi$).  Since each phase $\phi_{ij}=-\phi_{ji}$ is antisymmetric, the closure phase $\psi_{ijk}$ is totally antisymmetric; geometrically, it corresponds to the sum of the phases around an oriented triangle of baselines.  The closure phase is thus determined up to overall sign by the choice of three stations.  The (log) closure amplitude is constructed from a set of four stations $\{i,j,k,l\}$ by
\begin{align}\label{closure_amp}
    c_{ijkl} = \ln \left| \frac{V_{ij} V_{kl}} {V_{ik} V_{jl}} \right| = a_{ij} + a_{kl} - a_{ik} - a_{jl},
\end{align}
where $a_{ij}=\ln |V_{ij}|$ is the log-amplitude of each visibility.  Since $a_{ij}=a_{ji}$, the closure log-amplitude $c_{ijkl}$ has three independent index symmetries, $c_{ijkl}=c_{klij}=c_{jilk}=-c_{ikjl}$.  These imply that, given a choice of four stations, only $3$ of the $24$ closure log-amplitudes have numerically different absolute values.

Although the closure phase and amplitude are sometimes said to be \textit{immune} to station gain errors, this statement is not quite accurate in the context of model-fitting, where the full distribution (not merely the mean value) is required.  We will discuss the case of a single closure phase for simplicity; the full likelihood involving both closure quantities is precisely analogous (App.~\ref{app:like}).  In the limit of infinite signal to noise (no thermal noise at all), the closure phase is indeed independent of the residual station gains,
\begin{align}
    \psi_{ijk} = \arg (\mathcal{V}_{ij}\mathcal{V}_{jk}\mathcal{V}_{ki}), \qquad (\sigma=0).
\end{align}
This is of no use in model-fitting, however, since one must incorporate the thermal noise in order to construct a likelihood function. To see how this should be done, note that the error budget \eqref{visbudget} is equivalent to the statement that $V_{ij}$ is normally distributed with mean $g_i g^*_j \mathcal{V}_{ij}$ and variance $\sigma_{ij}^2$,
\begin{align}\label{visbudget2}
    V_{ij} \sim N(g_i g^*_j \mathcal{V}_{ij}, \sigma_{ij}^2).
\end{align}
In the high-SNR limit, any function of $V_{ij}$ will also be normally distributed, with (co-)variance determined by the usual propagation of errors. For a single closure phase defined by Eq.~\eqref{closure_phase}, this becomes
\begin{align}
\psi_{ijk} \sim N( \bar{\psi}_{ijk}, \textrm{var}(\psi_{ijk})),
\end{align}
where
\begin{align}
    \bar{\psi}_{ijk} & = \arg( \mathcal{V}_{ij}\mathcal{V}_{jk}\mathcal{V}_{ki} ) \label{mean} \\
    \textrm{var}(\psi_{ijk}) & = \frac{\sigma_{ij}^2}{|g_i g_j^* \mathcal{V}_{ij}|^2}+\frac{\sigma_{jk}^2}{|g_j g^*_k \mathcal{V}_{jk}|^2}+\frac{\sigma_{ki}^2}{|g_k g^*_i \mathcal{V}_{ki}|^2}. \label{variance}
\end{align}
Notice that the gain terms have canceled out of the mean value \eqref{mean}, but they remain present in the variance \eqref{variance}.  This means that the likelihood function $\mathcal{L}_{ijk}$ (the probability density for a measurement of the closure phase to be near some given value of $\psi_{ijk}$) will contain the gain terms as well.  Explicitly, the likelihood is given by
\begin{align}\label{likelihood}
    \mathcal{L}_{ijk} = \frac{1}{\sqrt{2\pi \textrm{var}(\psi_{ijk})}} \exp\left[-\frac{(\psi_{ijk}-\bar{\psi}_{ijk})^2}{2\textrm{var}(\psi_{ijk})} \right],
\end{align}
where $\bar{\psi}_{ijk}$ and $\textrm{var}(\psi_{ijk})$ are given by Eqs.~\eqref{mean} and \eqref{variance}, respectively.  To do model-fitting with closure phase we therefore still require the unknown residual gains to be part of the model parameters, and the closure phase approach does not, as it stands, resolve the issue it was meant to address.

We see two natural ways to proceed.  First, if the magnitude of the residual gain terms is expected to be reasonably near unity ($|g_i|\approx 1$ for all stations), then one could simply drop these terms in Eq.~\eqref{variance}, yielding
\begin{align}
    \textrm{var}(\psi_{ijk}) \approx \frac{\sigma_{ij}^2}{| \mathcal{V}_{ij}|^2}+\frac{\sigma_{jk}^2}{|\mathcal{V}_{jk}|^2}+\frac{\sigma_{ki}^2}{|\mathcal{V}_{ki}|^2}. \label{varianceA}
\end{align}
This would be appropriate for an array that had large phase calibration errors, but only more mild amplitude calibration errors.  Alternatively, if the SNR is very high ($|V_{ij}| \gg \sigma_{ij}$ for all involved baselines), one could argue that the observed value of $|V_{ij}|$ must be very close to its expected value $|g_i g_j^* \mathcal{V}_{ij}|$, and instead use
\begin{align}
    \textrm{var}(\psi_{ijk}) \approx \frac{\sigma_{ij}^2}{| \hat{V}_{ij}|^2}+\frac{\sigma_{jk}^2}{|\hat{V}_{jk}|^2}+\frac{\sigma_{ki}^2}{|\hat{V}_{ki}|^2}, \label{varianceB}
\end{align}
where $\hat{V}_{ij}$ are  the \textit{specific observed} values of the visibilities.  We are not aware of a limit in which this equation is strictly correct (the associated ``likelihood'' \eqref{likelihood} is never the probability density for $\psi_{ijk}$ predicted by the model \eqref{visbudget} or \eqref{visbudget2}), but it  nevertheless represents a reasonable choice for use in model-fitting when residual gains are large. 

In the context of fitting a model for the visibility $\mathcal{V}$, the two choices of likelihood differ in whether one uses the \textit{model} visibility $\mathcal{V}$ (Eq.~\eqref{varianceA}) or the \textit{data} visibility $\hat{V}$ (Eq.~\eqref{varianceB}) to compute the variance in Eq.~\eqref{likelihood}.  When the model is used (Eq.~\eqref{varianceA}), the variances depend on the model parameters, and we will refer to this as the ``variable'' likelihood.  When the data is used (Eq.~\eqref{varianceB}), the variances are fixed once and for all from the data, and we will refer to this as the ``fixed'' likelihood.  The variable likelihood is strictly correct for $|g|=1$ in the high-SNR regime, but can significantly misrepresent the actual likelihood when the residual gains are large.  The fixed likelihood is outside the framework of probabilistic model-fitting (either classical or Bayesian), but seems likely to perform well in the high-SNR regime already assumed in writing down Eq.~\eqref{likelihood}.  For an array like EHT, where large amplitude gains are possible and SNR is not always very high, we see no clear reason of principle to prefer one over the other, and both are problematic for low-SNR points.  

The EHT papers simply present the fixed likelihood as correct in the high-SNR limit, as does a standard text \citep[Eq.~10.55]{TMS}.  As far as we are aware, the variable likelihood has not been discussed previously, although analogous issues at moderate and low SNR were investigated in \citet{blackburn2020}.  In this paper we clarify the status of the fixed and variable likelihoods as approximations to the true high-SNR likelihood function.


\begin{figure*}
    \centering
    \includegraphics[width=0.39\textwidth]{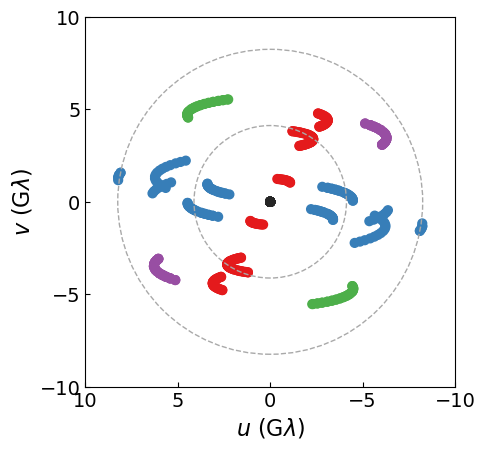}
    \hspace{5mm}
    \includegraphics[width=0.56\textwidth]{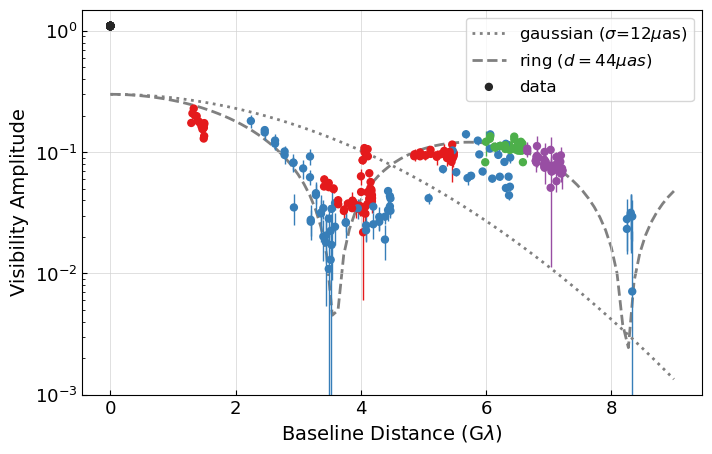}
     \caption{Visibilities of the April 06 hi-band dataset, displayed as in Fig.~1 of \citetalias{EHT6}. \textit{Left}: Fourier domain coverage of the array. As the Earth rotates, each baseline traces out a pair of arcs in the $u$--$v$ plane. Points are color-coded to highlight the presence of two main axes in the $u$--$v$ plane (red and blue), together with additional long-baseline coverage (purple and green) as well as intra-site baselines (black). The dotted lines represent the baseline distance needed to achieve a nominal resolution of $50\mu$as (inner circle) and $25\mu$as (outer circle). \textit{Right}: The corresponding scan-averaged visibility amplitudes, plotted as a function of baseline distance.  The visibility amplitudes of a Gaussian and an axisymmetric thin ring are shown for reference.  The intra-site baselines (black) probe flux at much larger scales than the Gaussian or ring are meant to represent. A dip near $3.5G\lambda$, and possibly another around $8G\lambda$, imply more structure than a simple Gaussian.}
    \label{fig:coverage-and-visamp}	
\end{figure*}

\section{EHT Data and Analysis}\label{sec:data}

\subsection{Data Processing}

The Event Horizon Telescope (EHT) is a global network of radio telescopes operating as an Earth-sized interferometer.  In 2017, the EHT observed M87* over four days in two similar frequency bands (called lo and hi), producing a total of 8 datasets. At the time, the EHT consisted of seven telescopes at five different geographical sites, resulting in $\binom{7}{2}$ = 21 total baselines.\footnote{An eighth station, the South Pole Telescope, was not used, since M87 is in the Northern hemisphere. Future observations will include additional stations.}  The two ``intra-site'' baselines aid in calibration and provide large-scale sky brightness information.  Of the remaining 19 baselines, $\binom{5}{2}$ = 10 provide independent $u-v$ coverage on the scales of interest.  As the Earth rotates, the projections of the EHT baselines onto the line of sight change, tracing out arcs in the $u$--$v$ plane (left panel of Figure~\ref{fig:coverage-and-visamp}).

The calibrated visibilites reported by EHT are grouped into ``scans'' of tens of data points over a few minutes' time, separated by longer pauses used for calibration.  Following EHTC, we analyze the data at the scan-averaged level.  That is, we take the mean of the baseline coordinates, timestamps, and complex visibilities and use the standard error (error on the mean) as the uncertainty on the visibility.  This reduces the size of each dataset to a few hundred points, depending on the observation day. 

The visibility amplitudes for one of these datasets are shown in Figure~\ref{fig:coverage-and-visamp} (right panel). As noted by EHTC, evidence for a ring-like structure is visible ``by eye'' from the periodic dips in visibility amplitude seen clearly at $4 G\lambda$, and less clearly at $8 G\lambda$.  Unfortunately, the residual station gains for the EHT array (unknown calibration factors in Eq.~\eqref{visbudget}) are significant enough that fitting a model to visibility amplitudes would require modeling gain terms as well.  As full gain modeling would be far more computationally intensive, EHTC considered two simpler alternatives.  The first used amplitudes and closure phases, including amplitude gain parameters but utilizing a Laplace approximation to marginalize over them instead of fully sampling the likelihood \citep{broderick-THEMIS}.  The second used closure amplitudes and closure phases without including gain parameters \citepalias{EHT6}.  In this paper we reproduce and extend the latter analysis, using closure phase and closure amplitude as our data products.

Following \citetalias{EHT6}, we will make two adjustments to the data before constructing likelihood functions.  First, we replace the measured (log-)amplitude with the \textit{debiased} (log-)amplitude,
\begin{align}\label{debiased}
    a_{ij} \rightarrow a_{ij}^{\rm deb.}, \qquad a_{ij}^{\rm deb.} = \ln \left( \sqrt{|V_{ij}|^2 - \sigma_{ij}^2} \right).
\end{align}
This correction arises from the idea that measured visibility amplitudes are biased upwards by the presence of thermal noise.  More precisely, the complex visibility $V_{ij}$ obeys a Gaussian distribution about some mean $\bar{V}_{ij}$, meaning that its amplitude $|V_{ij}|$ obeys a Rice distribution.  In the high-SNR limit the Rice distribution is approximately Gaussian with mean $|\bar{V}_{ij}|$, but at moderate SNR it is better approximated by a Gaussian about the larger mean value $(|\bar{V}_{ij}|^2 + \sigma_{ij}^2)^{1/2}$. Rather than using this up-adjusted mean value for the likelihood functions, we down-adjust the data via \eqref{debiased}.  Put another way, we work with the random variable given by (the log of) $(|V_{ij}|^2 - \sigma_{ij}^2)^{1/2}$, which is well-described by a Gaussian with mean equal to (the log of) $|\bar{V}_{ij}|$.

Amplitudes must be positive of course, so on the rare occasions that the measured amplitude is actually \textit{smaller} than the noise ($|V_{ij}|<\sigma_{ij}$), those data points are discarded. 

The second adjustment we make to the data, also following EHTC, is to add a small systematic error component to all visibilities. This is motivated in part by the fact that self-consistency tests on closure quantities suggest a small `non-closing' error \citepalias[Sec 8.4.5]{EHT3}.  We adopt the value of one percent used in \citetalias{EHT6}, increasing all variances in the data by
\begin{align}
 \sigma_{ij}^2 \rightarrow \sigma_{ij}^2 + \left( 0.01 |V_{ij}| \right) ^2.
\end{align}
This helps account for unknown systematics and sets an overall limit on the accuracy we can reasonably expect of the experiment.


\subsection{Likelihood Function}

To construct a likelihood function in terms of closure quantities, we must first select a set of \textit{linearly independent} closure quantities from the much larger set of \textit{all} such quantities that can be constructed from the stations in the array.  Although any such choice is formally valid from a statistical perspective, the best choice in practice is the one that ``squeezes'' the most SNR out of the data.  We will use the algorithm of \cite{blackburn2020}, which first sorts the candidate closure quantities by SNR and preferentially removes low-SNR quantities until a maximal set has been obtained.  Our specific construction is described in detail in App.~\ref{app:like} below.

However, as noted in Sec.~\ref{sec:closure} above, the correct high-SNR likelihood function still involves the unknown residual amplitude gain parameters for each station and each scan.  The issue is that the variance $\sigma_{ij}^2/|\bar{V}_{ij}|^2$ for each phase and log-amplitude involves the mean value $\bar{V}_{ij} = g_i g_j^* \mathcal{V}_{ij}$.  As we do not wish to include residual gain parameters in our model, some approximation must be made.  The two natural options are to replace the mean value $\bar{V}_{ij}$ with the observed value $\hat{V}_{ij}$ or the model mean value $\mathcal{V}_{ij}$.   When we use the observed value to construct the likelihood (Eq.~\eqref{variance_matrix} with $\bar{V}\to\hat{V}$), we call this a fixed likelihood, whereas when we use the model value  (Eq.~\eqref{variance_matrix} with $\bar{V}\to\mathcal{V}$) for some or all points, we call it a variable likelihood.

\begin{figure}
    \centering
    \includegraphics[width=0.95\linewidth]{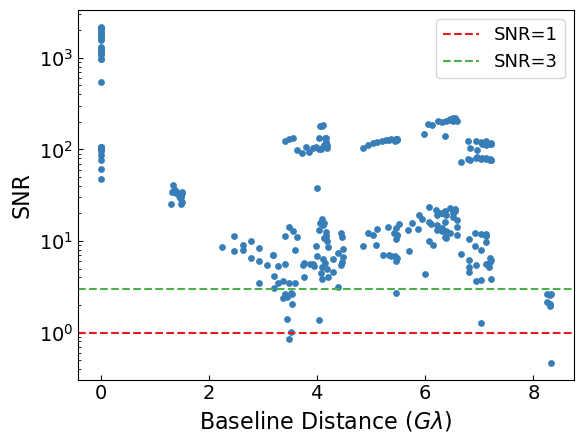} 
    \caption{Apr06-hi data plotted as a function of their SNR \eqref{SNR}. Several data points fall below the threshold of SNR=3, where the Gaussian likelihood approximation starts to break down.  Low-SNR points are concentrated near small visibility amplitudes, which are the very nulls that most strongly indicate a ring-like structure in the image.} 
    \label{fig:SNR}
\end{figure}

A second complication arises from the fact that the data are not entirely in the high-SNR regime, as our likelihood framework assumes.  Fig.~\eqref{fig:SNR} shows an example, where SNR is defined using the measured amplitudes $\hat{V}_{ij}$, 
\begin{align}\label{SNR}
    SNR_{ij} = \frac{|\hat{V}_{ij}|}{\sigma_{ij}}.
\end{align}
The low-SNR points naturally occur near the putative visibility minima around $3.8$G$\lambda$ and $8$G$\lambda$, whose properties would seem to be 
important for constraining the diameter and width of the ring.  Neither likelihood formulation is entirely valid for these low-SNR points.

The best approach for handling low-SNR points would be to go beyond the Gaussian approximation to include the full probability distribution for closure phase and closure log-amplitude.  While significant progress has been made \citep{christian-psaltis2020,blackburn2020}, this approach remains impractical, especially for closure amplitudes.  Restricting to the Gaussian approximation, the best one can do is to quantify whether these points are having an outsized influence on model posteriors  \citep{psaltis2020}.  For the fixed likelihood, we find that the low-SNR points (SNR<3.0) are not very important---the posteriors we derive are roughly independent of whether or not we include them.  For the variable likelihood, we encounter the difficulty that the variances $\sigma_{ij}^2/|\mathcal{V}_{ij}|^2$ can become arbitrarily large near nulls in the model visibility.  The Gaussian approximation is thus \textit{especially bad} for the variable likelihood near low-SNR points.  We can avoid the difficulty either by switching to the fixed likelihood ($\bar{V}_{ij}\to\hat{V}_{ij}$) for points below a given SNR cutoff, or by removing points below a given cutoff.  In this case we find that the different choices do affect the posteriors, and we study this dependence in detail.

Exploring the large space of reasonable (but strictly incorrect) likelihood functions discussed above, we have identified four choices that faithfully represent the range of variation in the results.  These are:
\begin{enumerate}
    \item $\bm{\mathcal{L}_\textrm{fix}}$: Use the fixed likelihood (the EHTC approach).
    \item $\bm{\mathcal{L}_{\textrm{var}}\!-\!3.0}$: Use the variable likelihood for visibilities with $|\mathcal{V}_{ij}|/\sigma_{ij}>3.0$; otherwise use the fixed likelihood.\footnote{For $\bm{\mathcal{L}_{\textrm{var}}\!-\!3.0}$ and $\bm{\mathcal{L}_{\textrm{var}}\!-\!0.1}$, we only make the switch to fixed likelihood if  $|\hat{V}_{ij}|>|\mathcal{V}_{ij}|$, i.e. if it actually improves the SNR.} 
    \item $\bm{\mathcal{L}_{\textrm{var}}\!-\!0.1}$: Use the variable likelihood for visibilities with $|\mathcal{V}_{ij}|/\sigma_{ij}>0.1$; otherwise use the fixed likelihood. (Essentially we always try to use the variable likelihood unless we hit a null in the model).
    \item $\bm{\mathcal{L}_{\textrm{var}}\!-\!\textrm{\textbf{remove}}}$: Discard the data points satisfying $|\hat{V}_{ij}|/\sigma_{ij}<3.0$ and use the variable likelihood.  
\end{enumerate}

Notice that choices (ii) and (iii) are ``mixed'' in that the variances \eqref{variance_matrix} are calculated with $\bar{V}\to\hat{V}$ for some visibilities and $\bar{V} \to \mathcal{V}$ for others, such that the final likelihood involve a mix of both.  Furthermore, the set of points being treated in each way changes depending on the value of the model parameters, such that the likelihood function is not the same at every step of the sampler.  This unusual situation is outside the confines of strict Bayesian analysis, but we have already breached this boundary by replacing the true likelihood with a fixed likelihood (\citetalias{EHT6}, reproduced here) or variable likelihood (this paper).  The fraction of points deemed low-SNR in the mixed likelihoods is always very low, and in the next section we will show that our results for variable likelihoods are largely independent of which method is used. 

Finally, note that variable likelihoods are more computationally costly than the fixed likelihood, as the covariance matrices ($\mathbf{\Sigma_\psi}$ and $\mathbf{\Sigma_c}$ in App.~\ref{app:like}) depend on the model and must be re-calculated with every sample. 

\section{Analysis of EHT data using the xs-ring model}\label{sec:analysis}

\subsection{The xs-ring Model}

The idea behind geometric modeling is to capture the gross features of possible images with a handful of parameters. These are not models in the astrophysical sense---there are no black holes or accretion disks---but merely ways of parameterizing a space of reasonable image shapes. An advantage of this approach is that one can work in a Bayesian framework and assign a likelihood to any given choice of model parameters, using a likelihood function such as the ones described above. This allows us to generate posterior probability distributions on the model parameters, quantifying the uncertainty on important features such as the diameter and width of the ring.  Of course, these posterior distributions are valid under the assumption that the true image is drawn from the space of model parameters, so a conclusion is only as robust as the model used to generate it.  In this kind of analysis, one wants a model that is flexible enough to cover a wide range of possibilities (so we aren't just imposing our beliefs about the data \textit{a priori}), but simple enough that these Bayesian likelihood methods can be applied.

We use the EHT collaboration's ``eccentric slashed ring'' model, or \textsf{xs-ring} for short.  This model is analytic in the Fourier domain (the complex visibilities are simple analytic functions of the model parameters), making it especially convenient for direct comparison  with the data.  It is also flexible enough to accommodate disks, rings, and crescents, with or without a brightness gradient, and we will add additional flexibility in the form of nuisance parameters.  The image domain appearance for a given set of parameters must be constructed numerically by inverse Fourier transform, but the broad features are easily understood from the model parameters.  We now describe these main features, leaving the details for Appendix~\ref{app:xs-ring} below.

The model \textsf{xs-ring} has 8 parameters, and is constructed as follows. We begin with a disk of uniform brightness and radius $R_\textrm{out}$. A smaller disk of radius $R_\textrm{in}$ is then subtracted, whose origin is shifted from the origin of the larger disk by a distance $r_0$ in the negative $x$-direction. This circular hole is given a non-zero uniform brightness characterized by $\gamma$, ranging from completely dark ($\gamma=0$) to no brightness depression at all ($\gamma=1$). A linear gradient of strength $\beta$ is then applied to the crescent (but not to the hole) in the x-direction, and the whole image is rotated by an angle $\phi$. Finally, the image is blurred with a Gaussian smoothing kernel of width $\sigma$. The whole crescent is normalized so that its total flux is determined by a single parameter, $V_0$. Favoring dimensionless parameters, the width of the crescent is characterized by $\psi = 1 - (R_\textrm{in}/R_\textrm{out})$, and the degree of asymmetry from $r_0$ is captured in $\tau = 1 - r_0/(R_\textrm{out}-R_\textrm{in})$. Precise definitions of all parameters, as well as the complete Fourier domain expression used to compute model visibilities, can be found in App.~\ref{app:xs-ring}. Given a choice of model parameters $\{ V_0, R_\textrm{out}, \phi, \psi, \tau, \beta, \gamma, \sigma \}$, a model prediction can be computed for each baseline $(u,v)$. 

One drawback of \textsf{xs-ring} is that, having only a handful of parameters, it is too simple to capture more complex features that might be present in the true sky brightness. A common approach in model fitting is to add what are called ``nuisance parameters'' -- additional degrees of freedom which do not necessarily represent actual features of the data, but allow a simple model to achieve a better fit. If the model posteriors are relatively unchanged by adding these nuisance parameters, so that their only effect is to improve the quality of the fit, then their inclusion helps to alleviate concerns that the posteriors cannot be trusted because of a relatively poor fit to the data. 

Following EHTC \citepalias{EHT6}, we will consider nuisance parameters that describe elliptical Gaussians in the image domain.  Each nuisance Gaussian adds 6 new parameters: two for the location of the origin ($x_0,y_0$), two for the width in each direction ($\sigma_x, \sigma_y$), one for the orientation angle ($\theta$), and one for the total flux ($V_g$) (see App.~\ref{app:xs-ring}
for details).  We explore a varying number of nuisance Gaussians in the results that follow.  

EHTC also included an additional large-scale Gaussian component to absorb unmodeled flux contributing only to the intra-site baselines of the array. Although our likelihood functions do include the intra-site baselines as part of the closure log-amplitudes, we have found that the presence of a large-scale component does not affect the derived-parameter posteriors. Since the addition of a poorly-constrained extra component makes sampling more difficult, we have not included it in the results presented in this paper.

\subsection{Priors and Sampling}\label{sec:priors-sampling}

The likelihood function depends on both the measured data and the model parameters.  At fixed model parameters (assuming the truth of the model at those specific model parameters), it represents the probability density for the data to be near the measured values.  Assigning prior probabilities to the parameter values allows this to be converted to a statement about the probability of the model parameters---the \textit{posterior probability}---via Bayes' theorem.  As our priors will be uniform, simply serving to enforce parameter ranges, the likelihood is proportional to the posterior probability density.   We will therefore view the likelihood as a function of model parameters, whose sampling provides the probability distribution function (PDF) for the  parameters.  When we discuss sampling the likelihood, we will mean the likelihood function multiplied by step functions enforcing parameter ranges.

The full list of priors used for model and nuisance parameters is given in Appendix~\ref{app:xs-ring}, but we mention one in particular here.  
Closure quantities have the property that the mean values are invariant under a rescaling of all visibilities, i.e., they are blind to the total flux density.  However, because we are exploring the \textit{variable} likelihood, in which the amplitudes of the model visibilities enter into the closure variances (e.g., Eq. \eqref{varianceA}), the overall scale of the flux does appear indirectly.  We found that this can cause headaches for the sampler, because  pathological `fits' to the data can arise in which very small visibility amplitudes result in very large error bars on the closure measurements.  To avoid these issues, we chose to fix the total flux in the ring at a constant value, i.e. impose a very strict prior on $V_0$.  We set $V_0 = 0.4$Jy in order to roughly fit the visibility amplitudes seen in Fig \ref{fig:coverage-and-visamp}.  In practice we find that sampling results are insensitive to the particular value chosen, as long as it is fixed.  We emphasize that we fix the flux in the ring component, not the total flux in the image.  The nuisance Gaussians are given freedom to explore any value of flux---the priors are very wide and the edges are not explored by the samplers.   We track the amount of flux in the nuisance Gaussians using a derived parameter $f_V$ defined below.

The dimensionality of the parameter space we sample ranges from 7 to 25, depending on the number of nuisance Gaussians included in the model (0--3).  To ensure that our results are independent of the sampler, we have used both a Markov-Chain Monte Carlo algorithm with the public code \textsf{EMCEE} \citep{emcee2013} and a nested sampling algorithm with the public code \textsf{DYNESTY} \citep{dynesty2020}.  As shown below, we find that the results agree exactly at low numbers of nuisance Gaussians, but begin to diverge slightly as the number of parameters increases.  We attribute this to minor issues with convergence and local minima, and use the discrepancy as part of our overall quantification of the uncertainty in our results.

\subsection{Derived Parameters}\label{sec:derived}

In our analysis we focus on four quantities $\{d,\theta,f_w,f_V\}$, derived from the raw parameters of \textsf{xs-ring}, that are most relevant to the questions we ask in this paper.  The precise definitions are given in App.~\ref{app:xs-ring} below.  For the ring-like regions of parameter space  favored by the data, the interpretations are 
\vspace{4pt}

\renewcommand{\arraystretch}{1.2}
\begin{tabular}{ |r|l| } 
    $d:$ & ring diameter \\ 
    $\theta:$ & position angle (East of North) of brightest part of  ring \\
    $f_w:$ & ring fractional width \\ 
    $f_V:$ & ring fractional flux density \\ 
\end{tabular}
\label{table:derived_params}
\vspace{4pt}

Our parameters $d$ and $\theta$ agree with those defined in \citetalias{EHT6}, where they were denoted $\hat{d}$ and $\hat{\phi}$, respectively.  Our fractional width parameter $f_w$ closely approximates the full-width at half-maximum (FWHM) of the ring divided by its diameter $d$, providing a better approximation than the analogous parameter $\hat{f}_w$ used in \citetalias{EHT6} (see discussion in App.~\ref{app:xs-ring} below).  Our fractional flux parameter $f_V$ is the flux density in the ring component divided by the total flux density, including any nuisance Gaussians.   (There is no corresponding parameter in the EHTC analysis). 

\begin{figure}
    \centering
    \includegraphics[width=\linewidth]{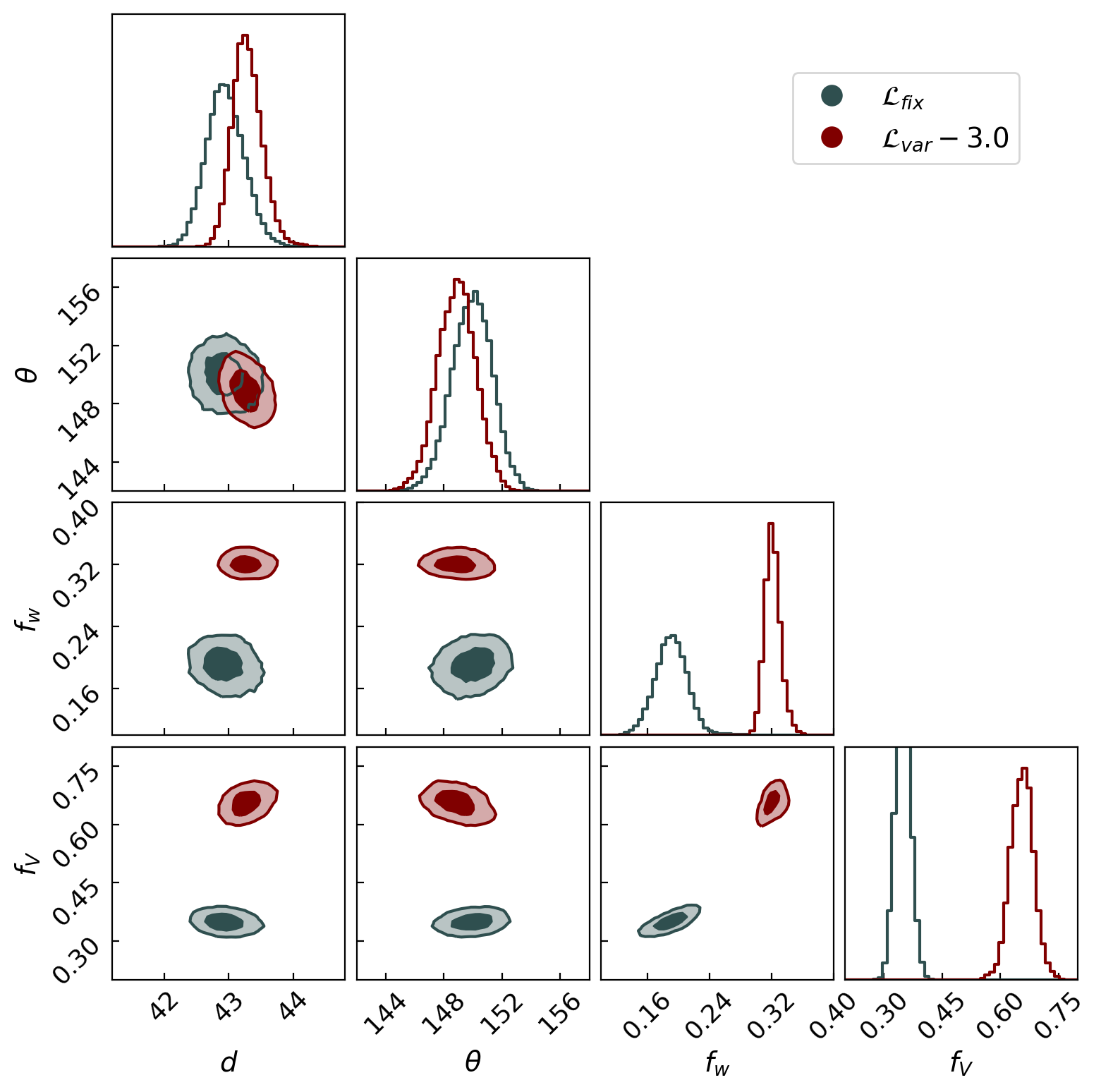} 
    \caption{Joint posterior probabilities for the derived parameters from the Apr 06 hi-band dataset, comparing fixed and variable likelihoods.  In each case we use 3 nuisance Gaussians and sample with \textsf{DYNESTY}.  Histograms and contours have been smoothed with a Gaussian kernel whose width is equal to half the width of one bin of the histogram (out of 50 bins), or $1\%$ of the width of the plots. The posteriors for the diameter $d$ and angle $\theta$ are overlapping, while the fractional width $f_w$ and flux $f_V$ are discrepant between likelihoods.}
    \label{fig:corner}
\end{figure}

\begin{figure}
    \centering
    \includegraphics[height=\linewidth]{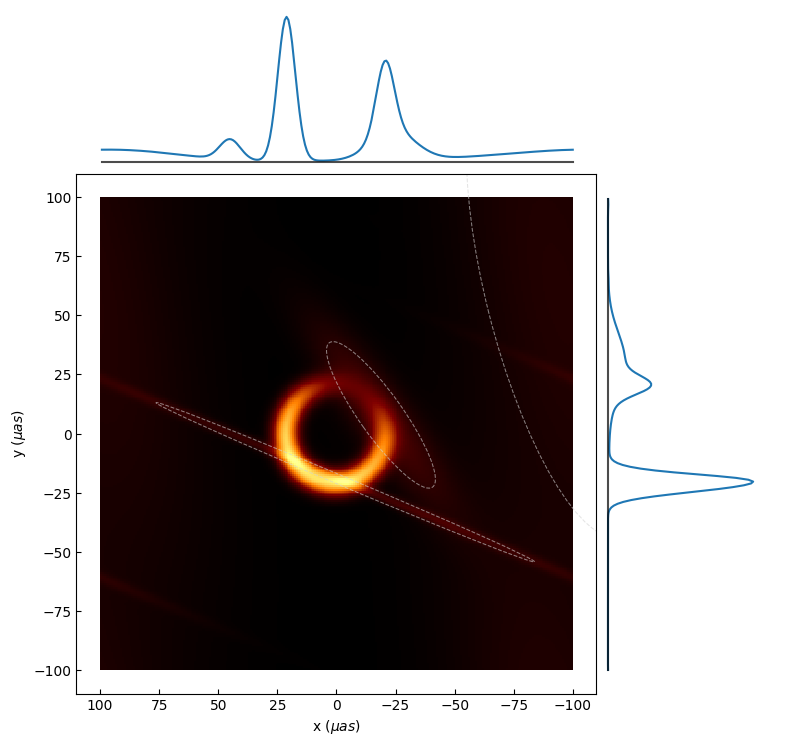}
    \includegraphics[height=\linewidth]{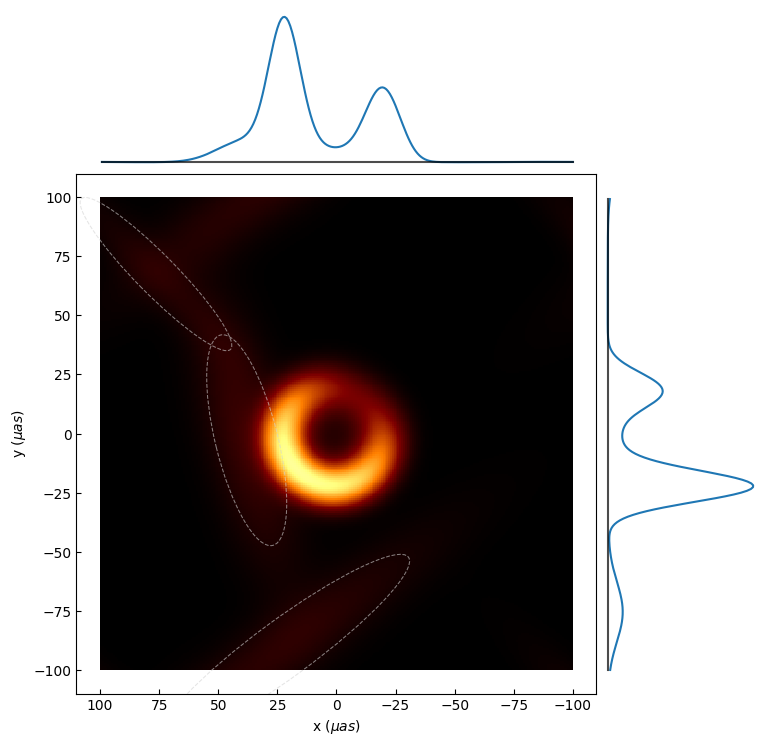}
    \caption{Image-domain appearance of the best-fitting parameters from the fixed-likelihood (top) and variable-likelihood (bottom) runs of   Fig.~\ref{fig:corner}.  These images are generated by numerically computing the inverse Fourier transform of the analytic visibility formulas in the \textsf{xs-ring} model, plotting the intensity in a linear color scale normalized to the brightest pixel, and adding dotted lines to outline the FWHM of the nuisance Gaussians.  Intensity cross-sections through $y=0$ and $x=0$ are plotted above and to the right of the images, respectively.  The fixed-likelihood best-fit (top) has derived parameters $d = 42.4\mu$as, $\theta = 148\degree$, $f_w = 0.20$, and $f_V = 0.37$.  The variable-likelihood best-fit (bottom) has parameters $d = 43.8\mu$as, $\theta = 147\degree$, $f_w = 0.31$, and $f_V = 0.66$. The difference in fractional width is evident by eye.  The nuisance Gaussians add structure to the ring and provide more diffuse intensity elsewhere.
    }
    \label{fig:images}	
\end{figure}

\subsection{Results}

As described above, for our \textsf{xs-ring} analysis we consider 8 datasets, 4 choices of likelihood, 4 choices of the number of nuisance Gaussians included, and 2 choices of sampler.  In this section we summarize the results of these 256 independent analyses, focusing on the derived parameters defined in Sec.~\ref{sec:derived} above.  We first show examples from the April 06 hi-band dataset (which has the greatest number of data points) before presenting topline results across datasets.

The derived-parameter posterior distributions from two example simulations are shown in Fig.~\ref{fig:corner}.  These runs use the same model and sampler (3 nuissance Gaussians and \textsf{DYNESTY}) on the same dataset (Apr 06 hi band), but differ in the choice of likelihood (fixed or variable).  The posterior distributions for the diameter $d$ and position angle $\theta$ are consistent between likelihood choices, but the fractional width $f_w$ and fractional flux $f_V$ are discrepant.  These features are evident in the best-fitting images for each likelihood, shown in Fig.~\ref{fig:images}.  The fixed likelihood result has a very narrow ring, while the variable likelihood result has a somewhat thicker ring.

\begin{figure*}
    \centering
    \includegraphics[width=0.95\textwidth]{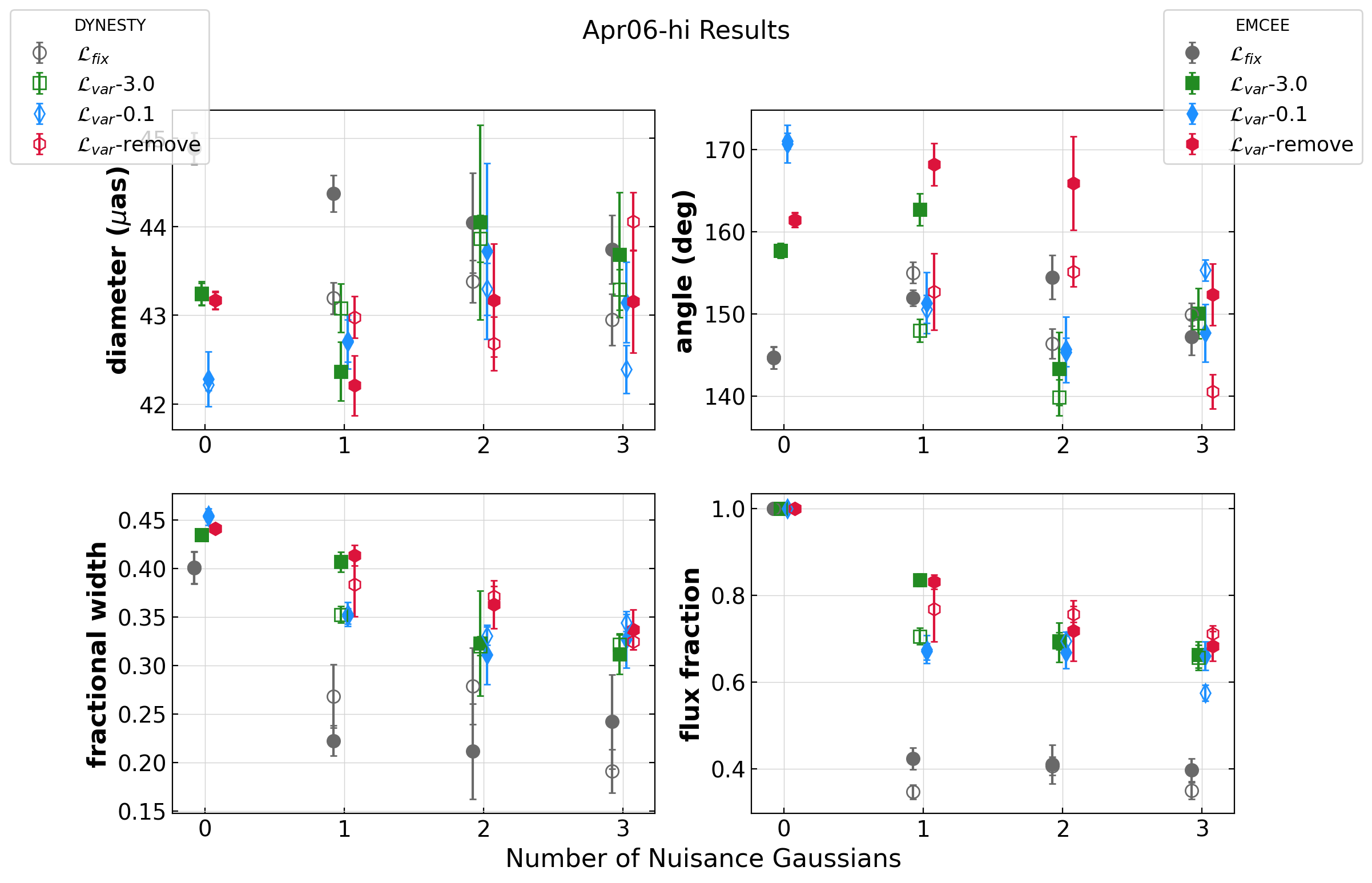}
    \caption{Results for the Apr 06 hi-band dataset.  We present the mean and standard deviations of four key parameters (Sec.~\ref{sec:derived}) across different choices of sampling algorithm, number of nuisance Gaussians, and likelihood function.}
    \label{fig:apr06hi_results}
\end{figure*}

The results for all 32 different analyses for this dataset  are summarized in Fig.~\ref{fig:apr06hi_results}.  For each derived parameter, we show the posterior mean and standard deviation inferred from each run (differing in choice of likelihood, sampler, and number of nuisance Gaussians).  For the most part, all runs with 2 or 3 nuisance Gaussians give consistent parameter values within each class of likelihood (fixed or variable).  However, there is a systematic discrepancy between fixed and variable for the fractional width and flux fraction, as already seen in the specific runs highlighted in Fig.~\ref{fig:corner}.

Based on these results, we conclude that the derived-parameter posteriors within each class of likelihood (fixed or variable) are sufficiently ``converged'' when there are 2 or 3 nuisance Gaussians.  To gain some kind of overall estimate of the favored parameter ranges of each class of likelihood, we take the mean values of all results from each likelihood class, with 2 or 3 nuisance Gaussians, and group them into a pseudo-dataset.  For the fixed likelihood the pseudo-dataset contains the 4 gray points on a given plot in Fig.~\ref{fig:corner} (two choices of sampler and two choices of the number of nuissance Gaussians), while for the variable likelihood it contains the 12 colored points (two samplers, two numbers of nuissance Gaussians, three likelihoods).  The mean value and standard deviation of these pseudo-datasets are reported in Fig.~\ref{fig:master_plot} above in the green-colored band, along with analogous results for the other 7 datasets.  

The systematic offset in fractional width and flux fraction between fixed and variable likelihoods is seen to persist across all 8 datasets.  We have emphasized that the fractional width is important for distinguishing among astronomical models.  The flux fraction, on the other hand, is largely of ``internal'' interest to the method, as it keeps track of how much flux is in the nuisance Gaussians, relative to the ring component, which is fixed at $V_0=0.4$Jy. While we caution against intepreting any particular configuration of nuisance Gaussians as real image features, we would note that the larger fractional fluxes of the variable likelihoods indicates that the nuisance Gaussians are doing ``less work'' to help the ring achieve a good fit.  This is consistent with the idea that the true, underlying sky appearance is a thicker ring than would be suggested by the fixed likelihood.

Figure~\ref{fig:master_plot} also helps understand any potential evolution of the derived parameters over the one-week observation window.  The diameter is very consistent at around $45 \mu$as.  The orientation angle changes relatively smoothly across the days, suggestive of physical changes in the source (rather than systematics of the experiment).  To interpret the position angle, we have added a reference line at $198 \degree$, which is the value that would be produced by Doppler boosting from matter orbiting in the plane perpendicular to the jet with negative angular momentum relative to the jet (see footnote \ref{foot:jet}).  We see no evidence for evolution of the fractional width, and the flux fraction is similarly stable.


\section{Outlook}\label{sec:discussion}

In this paper we have introduced a new approximation for the closure phase and closure amplitude likelihood function and explored its implications in the public M87* observations.  For the ring diameter and position angle, the new method gives results consistent with previous methods: there is an annular structure of approximately $40 \mu$as in diameter, with a mild brightness gradient that is brightest in the South to Southeast.  EHTC established these conclusions via agreement of three different image-domain methods and two different visibility-domain methods; to this list we add a third visibility-domain method reaching the same conclusions.

However, the new method gives different results for the fractional width of the ring.  The variable likelihood prefers ring widths of $30$--$40\%$, which are systematically larger than the fixed-likelihood widths of $10$--$25\%$.  While this is encouraging since the new method avoids the theoretically problematic regime of very narrow rings, it is also discouraging since we have no means, at present, to determine which approximation (fixed or variable) is more reliable for the EHT dataset.  All we can say with confidence is that the width is most likely somewhere in the range of $10$--$40\%$.


The state of affairs thus remains rather unsatisfactory, since there is a big difference in the theoretical interpretation in this range (App.~\ref{app:mechanisms}).  The low end would be very hard to explain theoretically, and thus potentially revolutionary; the low-middle regime seems to favor emission from near the horizon and would give some information about the black hole mass; the middle to upper regime is consistent with emission from further out, and from a black hole with a wide range of masses.  We hope that future papers in this series will help pin down the fractional width and begin to distinguish between these scenarios.


\section*{Acknowledgements}

In preparing this manuscript, we have benefited from numerous helpful interactions with EHT collaboration members.  We wish to particularly thank Lindy Blackburn, CK Chan,  Michael Johnson, Daniel Palumbo, Dom Pesce, Dan Marrone, and Maciek Wielgus for extensive correspondence about EHT methods, measurements, conventions, and claims.  We are further indebted to Dan Marrone for his patient instruction in the basics of radio VLBI.  Finally, we are grateful for useful comments provided by an anonymous referee. This work was supported in part by NSF grant PHY1752809 to the University of Arizona.


\section*{Data Availability}

The data analyzed in this study are described in \citetalias{EHT3} and available as \href{https://doi.org/10.25739/g85n-f134}{DOI:10.25739/g85n-f134}.  The conventions in this data are as follows: (1) The $uv$ coordinates are oriented such that $+v$ is North and $+u$ is East; (2) The complex visibility is constructed from the amplitude $A$ and phase $\varphi$ by $V=A e^{i \varphi}$; (3) The complex visibility is related to the sky brightness by $V=\int e^{+2\pi i (ux+vy)} I(x,y) dx dy$.  The latter formula differs by a sign in the exponential from the convention used here and in EHT papers (Eq.~\eqref{fourier_transform}), so the data must be complex-conjugated before being compared to formulas in this paper.



\bibliographystyle{mnras}
\bibliography{EHTdataI} 



\appendix

\section{Xs-ring model, priors, and derived parameters}\label{app:xs-ring}

In this appendix we review the \textsf{xs-ring} model \citepalias{EHT6}, describe our choice of priors, and define the derived parameters we focus on.

\subsection{Constructing the Crescent}

Here we describe the \textsf{xs-ring} model in detail. This construction is adapted from EHT VI Appendix B, and informed by the original slashed eccentric ring model of \cite{benkevitch2016}. 

We begin by considering one of the few shapes with an analytic 2D Fourier transform: a disk of uniform brightness. In the image domain the disk is defined as
\begin{align}
    D(r;R) = \begin{cases} 
     1, & r < R \\
     \, 0, & r > R
  \end{cases},
\end{align}
where $r=\sqrt{x^2+y^2}$ is the radial coordinate in the image domain.  We will denote Fourier transform (with the conventions of \eqref{fourier_transform}) with a tilde.  The Fourier domain representation of the disk is
\begin{align}
    \Tilde{D}(\rho;R) = \frac{R}{\rho} J_1 (2\pi \rho R),
\end{align}
where $\rho = \sqrt{u^2 + v^2}$ is the radial coordinate in the Fourier domain, and $J_1$ is a Bessel function of the first kind.

The crescent is constructed by taking a uniform disk of radius $R_\textrm{out}$ and subtracting a smaller disk of radius $R_\textrm{in}$ whose origin is shifted by a distance $r_0$ in the minus $x$-direction. A spatial shift in the image domain corresponds to a phase shift in the Fourier domain, and so the crescent is
\begin{align*}
   \Tilde{C}(u,v) &= \Tilde{D}_\textrm{out} - \mathrm{e}^{2\pi i r_0 u} \Tilde{D}_\textrm{in} \\
   &= \frac{R_\textrm{out}}{\rho} J_1 (2\pi \rho R_\textrm{out}) - \mathrm{e}^{2\pi i r_0 u} \frac{R_\textrm{in}}{\rho} J_1 (2\pi \rho R_\textrm{in}). 
\end{align*}

Next, a brightness gradient is applied across the crescent. The most natural way to do this would be to multiply the image by something like $(1 + \kappa\cos\theta)$ to create a dipole moment.\footnote{This is actually what you would expect to see for a uniformly radiating accretion disk that is viewed at an angle to the observer. The Doppler beaming effect is proportional to the line-of-sight velocities across the disk, which go as $\cos\theta$.}  This would involve Fourier-domain convolution with a Bessel function, which cannot be done in closed form.  Instead, we use what Benkevitch refers to as a ``slash'' operation, which applies a \textit{linear} gradient to the image. Let the minimum and maximum brightness at either end of the crescent be denoted $h_\textrm{min}$ and $h_\textrm{max}$. The slash operation is to multiply by
\begin{align}
    s_x = \bar{h} + \Delta h \left( \frac{x}{R_\textrm{out}} \right),
\end{align}
where $\bar{h} = (h_\textrm{max}+h_\textrm{min})/2$ is the brightness in the middle of the gradient, and $\Delta h = (h_\textrm{max}-h_\textrm{min})/2$. 
Note that this is equivalent to 
\begin{align}
    s_x = \bar{h} \left( 1 + \kappa \frac{x}{R_\textrm{out}} \right), \quad \kappa = \frac{h_\textrm{max}-h_\textrm{min}}{h_\textrm{max}+h_\textrm{min}},
\end{align}
meaning it is the same as multiplying by a dipole, except $\cos\theta = (x/r)$ is replaced with $(x/R)$. For thin rings, these have a very similar effect. The parameter we actually use to control the strength of the brightness gradient is $\ln \beta = \ln{ h_\textrm{max}/h_\textrm{min}}$.

Multiplication in the image domain corresponds to a derivative in the Fourier domain, so the slash gradient in Fourier space is
\begin{align}
    \Tilde{s}_u = \bar{h} + \frac{\Delta h}{R_\textrm{out}} \frac{i}{2\pi} \frac{\partial}{\partial u}.
\end{align}

After the gradient has been applied, a central ``floor'' is added, giving the hole a brightness that can range from zero to the full brightness of the crescent. This floor is just
\begin{align}
    \Tilde{F} = K \Tilde{D}_\textrm{in} = K \mathrm{e}^{2\pi i r_0 u} \frac{R_\textrm{in}}{\rho} J_1 (2\pi \rho R_\textrm{in}). 
\end{align}
The slashed crescent up to this point is therefore
\begin{align}
    \Tilde{s}_u[\Tilde{D}_\textrm{out} - \Tilde{D}_\textrm{in}] + \Tilde{F}.
\end{align}
The entire crescent is then rotated by an angle $\phi$. This is achieved by replacing $(u,v) \rightarrow (u\cos\phi + v\sin\phi, -u\sin\phi + v\cos\phi)$. Then the image is blurred by multiplication with a Gaussian of width $\sigma$. 
Finally, the flux is scaled via a geometric factor $\Gamma$ so the total flux is unity, and then multiplied by the desired flux parameter $V_0$. The final xs-ring formula in the Fourier domain is\footnote{Note that Eq. (41) of \citetalias{EHT6} contains a trivial typo, writing $R$ for $R_{\rm out}$.}
\begin{align}
\boxed{
   \mathbf{xs\text{-}ring}(u,v) = \frac{V_0}{\Gamma} \, \mathrm{e}^{-2\pi \sigma^2 \rho^2} \left( \Tilde{\mathcal{D}}_\textrm{out} - \mathrm{e}^{2\pi i r_0 u'} (\Tilde{\mathcal{D}}_\textrm{in} - \Tilde{F}) \right) 
}
\end{align}
where the slashed disks are given by
\begin{align*}
     \Tilde{\mathcal{D}}_\textrm{out} &= \frac{R_\textrm{out} (1+\beta)}{2\rho} J_1(X_\textrm{out}) - i \left( \frac{1-\beta}{ 4\pi \rho^2} \right) \\ 
     &\times \left( \pi R_\textrm{out} [J_0(X_\textrm{out}) - J_2(X_\textrm{out})] - \frac{1}{\rho} J_1(X_\textrm{out}) \right) u' ,
\end{align*}
\begin{align*}
    \Tilde{\mathcal{D}}_\textrm{in} &= \frac{R_\textrm{in}}{2\rho} \left( (1+\beta) + \frac{r_0}{R_\textrm{out}} (1-\beta) \right) J_1(X_\textrm{in}) - i \left( \frac{1-\beta}{4\pi \rho^2} \right) \frac{R_\textrm{in}}{R_\textrm{out}} \\
    &\times \left(\pi R_\textrm{in} [J_0(X_\textrm{in}) - J_2(X_\textrm{in})] - \frac{1}{\rho} J_1(X_\textrm{in}) \right) u' ,
\end{align*}
and we define
\begin{align*}
    u' &= u\cos\phi + v\sin\phi \\
    X_\textrm{out} &= 2 \pi \rho R_\textrm{out} \\
    X_\textrm{in}  &= 2 \pi \rho R_\textrm{in} 
\end{align*}
and
\begin{align*}
    \Tilde{F} &= \gamma \beta \frac{R_\textrm{in}}{\rho} J_1(X_\textrm{in}) \\
    \Gamma &= \frac{1}{2} \pi R_\textrm{out}^2 \left( (1+\beta) - (1-\psi)^2 [(1+\beta) - \psi (1-\tau) (1-\beta) - 2\gamma\beta] \right) \\
    \psi &= 1 - \frac{R_\textrm{in}}{R_\textrm{out}} \\
    \tau &= 1 - \frac{r_0}{R_\textrm{out}-R_\textrm{in}}. \\
\end{align*}

The parameterization used in the code are the eight parameters $\{ V_0, R_\textrm{out}, \phi, \psi, \tau, \ln \beta, \gamma, \sigma \}$.  The dimensionless parameter $\psi$ is used in place of $R_\textrm{in}$, and controls the width of the ring. $\gamma$ ranges from 0 to 1 and controls the relative brightness of the central floor. The parameter $\tau$ is used in place of $r_0$, and controls the degree of asymmetry of the crescent shape. 

In addition to the \textsf{xs-ring} model itself, EHTC added a much broader, large-scale Gaussian component to capture flux on the scales to which the intra-site baselines are sensitive. This large-scale Gaussian had a total flux constrained to lie between $0$ and $10 Jy$, and a width $10^{-2} < \sigma_G < 10^1$ arcsec. We found that when this component was included, it was highly unconstrained and did not affect the posteriors for our derived parameters. For this reason we omitted it from our modeling. 

\subsection{Nuisance Gaussians}\label{sec:nuisance}

Our nuisance parameters take the form of elliptical Gaussians. In the image domain these are 
\begin{align}
    G(x,y) = \frac{V_g}{2\pi \sigma_x \sigma_y} \mathrm{e}^{- \left[ a(x-x_0)^2 + 2b(x-x_0)(y-y_0) + c(y-y_0)^2 \right]},
\end{align}
with 
\begin{align*}
    a &= \frac{\cos^2 \theta}{2 \sigma_x^2} + \frac{\sin^2 \theta}{2 \sigma_y^2} \\
    b &= -\frac{\sin 2\theta}{4 \sigma_x^2} + \frac{\sin 2\theta}{4 \sigma_y^2} \\
    c &= \frac{\sin^2 \theta}{2 \sigma_x^2} + \frac{\cos^2 \theta}{2 \sigma_y^2}. 
\end{align*}
The Gaussian is centered at $(x_0,y_0)$ with width parameters $\sigma_x$ and $\sigma_y$ corresponding (respectively) to the $x$ and $y$ axes after a counter-clockwise rotation by $\theta$.  This parameterization is convenient because the parameter degeneracies are completely removed by a restriction $0<\theta<\pi$ on the angle $\theta$.  The Fourier transform (complex Visibility) is\footnote{This is a correction to the formula appearing in EHT VI eq. 47, which is missing a minus sign in the first exponential.}
\begin{align}
\begin{split}
    \Tilde{G}(x,y) &= V_g \, \mathrm{e}^{- 2\pi i (u x_0 + v y_0)} \, \mathrm{e}^{-4\pi^2 \sigma_x^2 \sigma_y^2 (c u^2 - 2 b u v + a v^2)}.
\end{split}
\end{align}

When there are at least two Gaussians, we must also remove degeneracies associated with swapping two or more of them.  We label the Gaussians with an index $i$ and parameterize the first Gaussian with $x_0$ and $y_0$, after which we instead consider the differences $\Delta x_0$ and $\Delta y_0$ from the position of the previous Gaussian,
\begin{align}
    \Delta {x_0}_{i+1} & = {x_0}_{i+1} - {x_0}_i \\
    \Delta {y_0}_{i+1} &= {y_0}_{i+1} - {x_0}_i.
\end{align}
We may then remove the degeneracy by demanding that $\Delta x_0$ be positive.  Our parameterization of nuisance Gaussians is identical to that used for $\textsf{xs-ring}$ in  \citetalias{EHT6}.

\subsection{Priors}

\begin{table}
\vspace{10pt}
\begin{tabular}{ |c|c|c| } 
    \hline
    & Parameter & Prior Range \\
    \hline
    \multirow{8}{1em}{\rotatebox[origin=c]{90}{xs-ring}} & $V_0$ & 0.4 Jy (fixed) \\ 
    & $R_\textrm{out}$ & [0, 50] $\mu$as \\ 
    & $\phi$ & [0, 2$\pi$] \\
    & $\psi$ & [0, 1] \\ 
    & $\tau$ & [0.5, 1] \\ 
    & $\ln \beta$ & [-5, 5] \\ 
    & $\gamma$ & [0, 1] \\ 
    & $\sigma$ & [0, 20] $\mu$as \\
    \hline
    \multirow{8}{1em}{\rotatebox[origin=c]{90}{nuisance}} & $V_g$ & [0, 3] Jy \\
    & $x_0$ & [-100, 100] $\mu$as \\
    & $y_0$ & [-100, 100] $\mu$as \\
    & $\Delta x$ & [0, 100] $\mu$as \\
    & $\Delta y$ & [-100, 100] $\mu$as \\
    & $\sigma_x$ & [0, 100] $\mu$as \\
    & $\sigma_y$ & [0, 100] $\mu$as \\
    & $\theta$ & [0, $\pi/2$] \\
    \hline
\end{tabular}
\caption{Priors ranges on all model parameters. All priors are uniform.}
\label{table:priors}
\vspace{10pt}
\end{table}

\begin{figure}
    \centering
    \includegraphics[width=\linewidth]{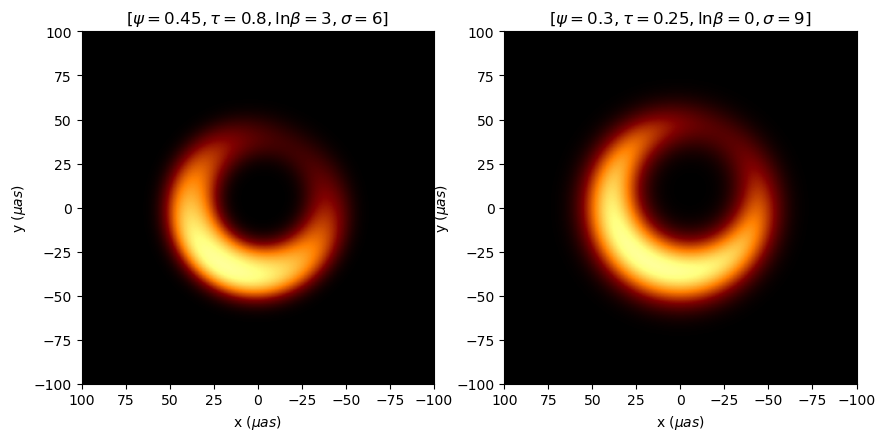}
    \caption{An example of a near-degeneracy in \textsf{xs-ring} parameters. When combined with the freedom in ring width ($\psi$) and blur 
    ($\sigma$), either a steep brightness gradient (left) or a large crescent asymmetry (right) can produce very similar images.}
    \label{fig:degeneracy}
\end{figure}

Table \ref{table:priors} lists the priors on each parameter used for model fitting. These priors are identical to those used in \citetalias{EHT6}, with two exceptions.  First, we impose a strict prior on $V_0$, as discussed in Sec.~\ref{sec:priors-sampling} above.  Second, we constrain the parameter $\tau$ to lie in the range [0.5,1] instead of [0,1].  This excludes highly asymmetric crescent shapes that we have found to be excluded anyway in sampling with the less restrictive prior, while fixing a parameter degeneracy illustrated in Fig.~\ref{fig:degeneracy}.  That is, in sampling with the less restrictive prior, we found that the preferred observational appearance was always ring-like, but the underlying parameter distribution was strongly bimodal, indicating an important parameter degeneracy.  In experiments with fewer nuisance Gaussians where we were able to reliably sample this bimodal distribution, we confirmed that the derived-parameter posteriors were unaffected if we eliminated the degeneracy with a more restrictive prior on $\tau$.  We then imposed this prior on all runs,  facilitating the robust inference of derived-parameter posteriors.

Apart from these two exceptions, all priors either enforce physical parameter ranges or give the parameters enough ``room'' such that the edge of the parameter range is never explored by the sampler.  Note that the uniform prior is on $\ln \beta$, rather than $\beta$, in order to be agnostic as to the direction of the brightest part of the ring.  As discussed in Sec.~\ref{sec:nuisance} above, the ranges on $\theta$ and $\Delta x$ remove true degeneracies in the parameterization of the Gaussians.

\subsection{Derived Parameters}

Here we define the four key parameters $\{d,\theta,f_w,f_V\}$ used in the analysis above. The diameter $d$ is defined as
\begin{align}
    d \equiv R_\textrm{out}+R_\textrm{in}
\end{align}
and describes the overall size of the crescent. This is precisely the mean diameter when the ring is symmetric ($\tau=1$).

The parameters $\beta$ and $\phi$ encode the direction of the brightness gradient.  The brightest part of the ring is at the angle $\phi$ if $\ln \beta>0$ and at the opposite location $\phi+\pi$ if $\ln \beta <0$.  Here $\phi$ is the usual polar angle, meaning the angle North of East if $+x$ is East and $+y$ is North.  To measure the angle East of North, we instead want to measure the angle $y$-axis in the opposite sense.  The angle East of North to the brightest part of the ring thus is
\begin{align}\label{theta}
    \theta \equiv \textrm{sign}(\ln \beta) \cdot 90\degree - \phi
\end{align}
modulo $2\pi$.  We consistently adopt the convention that $+x$ is East and $+y$ is North.  However, when plotting images we will have $x$ increase to the left (simulating looking up at the sky), such that the brightest portion appears at a counter-clockwise rotation by $\theta$ from the vertical, looking down on the page.\footnote{\citetalias{EHT6} defines $\hat{\phi}$ to be the angle East of North to the brightest part of the ring.  This makes it equal to our $\theta$, and indeed our posterior ranges are consistent under the identification $\theta=\hat{\phi}$.  However, the paper also incorrectly states that $\hat{\phi}=\phi$; the correct formula is Eq.~\eqref{theta} with $\theta=\hat{\phi}$.}

\begin{figure}
\centering
\includegraphics[width=\linewidth]{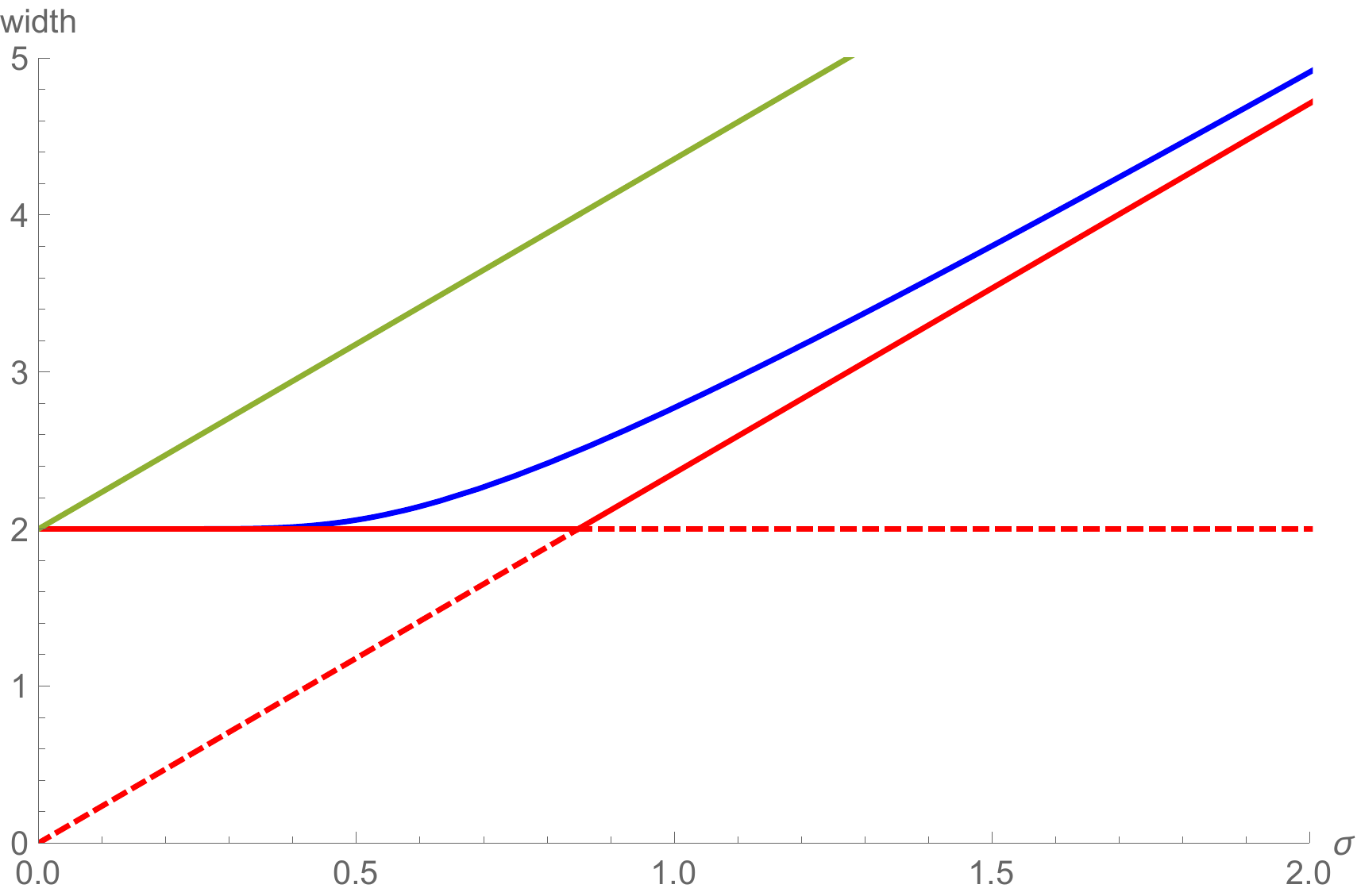} 
\caption{Approximations for the FWHM of a one-dimensional tophat of radius $1$ (width $2$) convolved with a Gaussian of standard deviation $\sigma$.  The numerically computed FWHM (blue) is compard to that of the tophat (red horizontal line) and the Gaussian (slanted red line through the origin).  The maximum of the two (solid red) is a better approximation than the sum (solid green).}
\label{fig:FWHM}
\end{figure}

The fractional width is defined as 
\begin{align}\label{fw}
    f_w \equiv \max \left( \frac{R_\textrm{out}-R_\textrm{in}}{d} \, , \, \frac{\sigma^*}{d} \right), \qquad \sigma^* = 2 \sigma \sqrt{2\ln(2)},
\end{align}
which closely approximates the FWHM of a ring (Fig.~\ref{fig:FWHM}) divided by its diameter.  The approximation is constructed by taking the maximum of the FWHM of a step-function ring (appropriate for small $\sigma$) and the FWHM of a Gaussian ring (appropriate for large $\sigma$).  EHTC instead used the sum of these two terms, which generally overestimates the FWHM (Fig.~\ref{fig:FWHM}).  When $\tau<1$, the crescent is thicker on one side than the other, and \eqref{fw} represents a typical fractional width.  This distinction is unimportant in the parameter regime $\tau \approx .8$ favored by the data.

Finally, we look at the ratio of flux in the ring to the total flux (ring + Gaussians), denoted
\begin{align}
  f_V \equiv \frac{V_0}{V_0 + \sum_i V_{g_i}}, 
\end{align}
where $V_g$ are the flux from each Gaussian. This tells us how much flux is attributed to the nuisance parameters compared to the ring. 

\section{Full likelihood description}\label{app:like}

In this appendix we go through the algorithm used to select an optimal set of closure phases and closure amplitudes, and describe the construction of the associated covariance matrices and likelihood functions.  We begin with a convenient mathematical description, following \cite{blackburn2020}.  An $N$-element interferometer produces $n=N(N-1)/2$ complex visibilities per observation.  Adopting a canonical ordering $1,2,\dots N$ of the stations, we may represent these visibilities as two $n$-dimensional column vectors containing all the phases and log-amplitudes in canonical order,
\begin{align}
    \Phi & = (\phi_{12},\phi_{13}, \dots, \phi_{1N}, \phi_{23}, \phi_{24}, \dots, \phi_{2N}, \dots, \phi_{(N-1)N} )^\top \\
    A & = (a_{12},a_{13}, \dots, a_{1N}, a_{23}, a_{24}, \dots, a_{2N}, \dots, a_{(N-1)N} )^\top.
\end{align}
The closure phases and closure log-amplitudes are linear functions of $\Phi$ and $A$ (respectively), meaning that they may be represented as row vectors.\footnote{This linearity makes the log-closure amplitude more convenient than the ordinary closure amplitude.}  The closure phase row vectors contain precisely three non-zero elements, each equal to $+1$ or $-1$, such that a valid closing sum \eqref{closure_phase} is formed when multiplied by (dotted with) the column-vector $\Phi$.  Similarly, the closure amplitude row vectors contain precisely four non-zero elements, each equal to $+1$ or $-1$, such that a valid closing sum \eqref{closure_amp} is formed when multiplied by (dotted with) the column-vector $A$.

A choice of $m_\psi$ closure phases may be represented as the $m_\psi \times n$ matrix of closure-phase row vectors, the \textit{design matrix} $\bm{D}_\psi$ for that set.  The rank $N_\psi$ of the design matrix is the number of linearly independent closure phases.  If $m_\psi > N_\psi$ then the closure phases are linearly dependent and some must be removed.  The idea of \cite{blackburn2020} is to preferentially remove the lowest-SNR closure phases.  We define the SNR as
\begin{align}\label{closure_phase_SNR}
    \frac{1}{SNR_{\psi_{ijk}}} \equiv \sqrt{
    \frac{\sigma_{ij}^2}{|\hat{V}_{ij}|^2} + \frac{\sigma_{jk}^2}{|\hat{V}_{jk}|^2} + \frac{\sigma_{ki}^2}{|\hat{V}_{kj}|^2} }, 
\end{align}
where $|\hat{V}_{ij}|$ is the reported raw visibility amplitude (i.e., not debiased).  We order the rows in the design matrix by SNR, starting with the lowest SNR.  Then we proceed down the list and for each row, eliminate it, as long as the resulting matrix still has the same rank. In other words, we remove the lowest-SNR closure phases as long as we can do so without reducing the dimension of the space.

We use the same algorthm for closure log-amplitudes, defining the SNR analogously as 
\begin{align}\label{closure_amp_SNR}
    \frac{1}{SNR_{c_{ijkl}}} &\equiv \sqrt{ 
    \frac{\sigma_{ij}^2}{|\hat{V}_{ij}|^2} + \frac{\sigma_{kl}^2}{|\hat{V}_{kl}|^2} + \frac{\sigma_{ik}^2}{|\hat{V}_{ik}|^2} +
    \frac{\sigma_{jl}^2}{|\hat{V}_{jl}|^2} }.
\end{align}
That is, given a choice of $m_c$ closure amplitudes represented as an $m_c \times n$ design matrix $\bm{D}_c$, we sort the rows by SNR and remove rows, from the top down, which do not reduce the rank $N_c$.

We implement this algorithm for each scan of each dataset, separately for closure amplitude and closure phase.  For the closure phase, we begin by forming all 3-element subsets (triangles) of the active stations in the scan.\footnote{Stations may be inactive during a scan because of observing issues or because the source is not visible on their sky.}  We then remove any ``trivial triangles'' that contain intrasite baselines, as these have near-zero closure phase.  To each remaining triangle we associate a closure phase using a canonical ordering of the stations.  For example, for stations $2,3,5$ we use
\begin{align}
    \psi_{235} = \phi_{23} + \phi_{35} - \phi_{25}.
\end{align}
This provides a list of closure phases with the 6-fold index-permutation degeneracy removed.  This set is then pared down further using the SNR-maximizing algorithm described above, resulting in a final design matrix $\bm{D}_\psi$ whose row-dimension is equal to its rank $N_\psi$, i.e., a maximal set of closure phases.

For the closure log-amplitude, we begin by forming all 4-element subsets of the active stations in the scan.  Following EHTC, we do \textit{not} remove the log-amplitudes including intrasite baselines.  As discussed in Sec.~\ref{sec:closure} above, to each 4-element subset there are associated three closure log-amplitudes whose absolute values are numerically different.  For each 4-element subset we make the following canonical choices,
\begin{align}\label{closure_amp_triplet}
\begin{split}
    c_{1234} &= a_{12} + a_{34} - a_{13} - a_{24} \\
    c_{1243} &= a_{12} + a_{34} - a_{14} - a_{23} \\
    c_{1342} &= a_{13} + a_{24} - a_{14} - a_{23},
\end{split}
\end{align}
where $1234$ stands for the station indices of the subset, sorted in ascending numerical order.  This list is then reduced to a valid maximal set by the SNR-maximizing algorithm described above, resulting in a final design matrix $\bm{D}_c$ whose row-dimension is equal to its rank $N_c$.

For each scan, the values of the closure phases and closure amplitudes of interest are constructed from the amplitudes and phases by matrix multiplication,  
\begin{align}\label{matrix_product}
    \Psi & = \mathbf{D}_\psi \, \Phi \\
    C & = \mathbf{D}_c \, A.
\end{align}
Here $\psi$ is the $N_\psi$-dimensional column-vector of closure phases, while $C$ is the $N_c$-dimensional column-vector of closure log-amplitudes. 

In the high-SNR limit we assume, each visibility phase $\phi_{ij}$ and log-amplitude $a_{ij}$ has variance given by
\begin{align}
    \textrm{var}(\phi_{ij}) = \textrm{var}(a_{ij}) =\frac{\sigma_{ij}^2}{|\bar{V}_{ij}|^2},
\end{align}
where $\bar{V}_{ij}$ is the mean value of the random variable $V_{ij}$.  (Recall from Sec.~\ref{sec:closure} above that this quantity contains residual gain terms, and in practice we replace $|\bar{V}_{ij}|$ with either the model mean value $|\mathcal{V}_{ij}|$ or the data mean value $|\hat{V}_{ij}|$.)  Since the uncertainty on the phase and log-amplitude takes the same value (and since the phases and log-amplitudes are all statistically independent), we may capture the full covariance of the scan data with a single $n\times n$ matrix $\bm{S}$ defined by
\begin{align}\label{variance_matrix}
    \mathbf{S} = \mathrm{diag}\left(\frac{\sigma^2_{12}}{|\bar{V}_{12}|^2},\frac{\sigma^2_{13}}{|\bar{V}_{13}|^2},...,\frac{\sigma^2_{(N-1)N}}{|\bar{V}_{(N-1)N}|^2}\right).
\end{align}
The covariance matrices for the closure phases and closure log-amplitudes are then given simply by
\begin{align}\label{cov_matrix}
    \mathbf{\Sigma_\psi} & = \mathbf{D}_\psi \, \mathbf{S} \, \mathbf{D}_\psi^\top, \\
    \mathbf{\Sigma_c} & = \mathbf{D}_c \, \mathbf{S} \, \mathbf{D}_c^\top.
\end{align}
The closure phase and closure log-amplitude likelihood functions for the scan are then given by
\begin{align}\label{likelihoods}
    \mathcal{L}_\psi & = \frac{1}{\sqrt{(2\pi)^{N_\psi} \det \mathbf{\Sigma_\psi} }} \exp\left[-\frac{1}{2} \Delta\!\Psi^\top \mathbf{\Sigma_\psi}^{-1} \Delta\!\Psi \right]. \\
    \mathcal{L}_c & = \frac{1}{\sqrt{(2\pi)^{N_c} \det \mathbf{\Sigma_c} }} \exp\left[-\frac{1}{2} \Delta\!C^\top \mathbf{\Sigma_c}^{-1} \Delta\!C \right],
\end{align}
where $\Delta \Psi$ and $\Delta C$ are $n$-dimensional column-vectors of closure-phase and closure-log-amplitude residuals (data minus model), respectively.

The final likelihood for a given dataset is the product of the $\mathcal{L}_\psi$ and $\mathcal{L}_c$ for all scans in the dataset.  In practice, of course, it is easier to work with the log-likelihood, which is the sum of all the individual log-likelihoods.  We further find it convenient to perform the sum-over-scans at the level of linear algebra by constructing block-diagonal design matrices $\bm{D}_\psi^{\rm tot}$ and $\bm{D}_c^{\rm tot}$ which contain all the individual design matrices for each scan.  We similarly combine the closure phases into a single vector $\Delta \Phi^{\rm tot}$, combine the closure amplitudes into a single vector $\Delta C^{\rm tot}$, and combine the uncertainties diagonal matrix into a single diagonal matrix $\bm{S}^{\rm tot}$.  This allows the final likelihood for the dataset to be constructed from just a few matrix operations.


\section{Goodness of Fit}\label{app:goodness}

In classical statistical modeling the measured quantities are considered independent Gaussian random variables with known variances but unknown mean values.  The quality of a model fit for the mean values can be assessed by seeing whether the data scatter about the model mean values with the known variances, e.g. by using a Kolmogorov–Smirnov test on the normalized residuals.  Alternatively, one can compute the $p$-value of the best fitting model, defined as the probability that randomly-generated mock data (drawn from Gaussian distributions using the best-fit mean values and known variances) has a lower likelihood than the measured data.  If this $p$-value is near $0$, then the mock data rarely reproduce the qualitative properties of the actual data and the fit is considered poor.  If the $p$-value is near $1$, then mock  data never do as well as the real data and one worries about an ``overfit'' to the specific noise in the experimental realization.  Intermediate $p$-values indicate an acceptable fit.  

Neither the residual test nor the $p$-value test is satisfactory for the modeling in this paper.  The closure quantities  are non-linear, covariant functions of the underlying complex visibilities, and the likelihood is not treated consistently because the residual gain terms are dropped (either using the data mean value or the model mean value in the formula for the variance).  Some data points are outside the Gaussian (high-SNR) regime, and the manner in which we treat these points does affect the results. 
Finally---and perhaps most importantly---the model is clearly not flexible enough to accommodate the actual detailed image appearance, instead using nuisance Gaussians to absorb unmodeled features.  In summary, there is no reason to expect a ``good fit'', and (correspondingly) no way to reasonably judge whether one has occurred.  

In such a situation, we avoid all putative measures of absolute fit quality and instead focus on \textit{relative} measures among different choices we make.  We want to know that the different choices all lead to reasonably similar fit quality, and we want to know that fit quality improves as we add parameters in the form of nuisance Gaussians.  If the fit quality is comparable among arbitrary choices and improves as we add nuisance parameters, and if the posteriors for the parameters of interest are relatively stable among the choices, then we declare success and present results for those parameters.

\begin{figure}
    \centering
    \includegraphics[width=\linewidth]{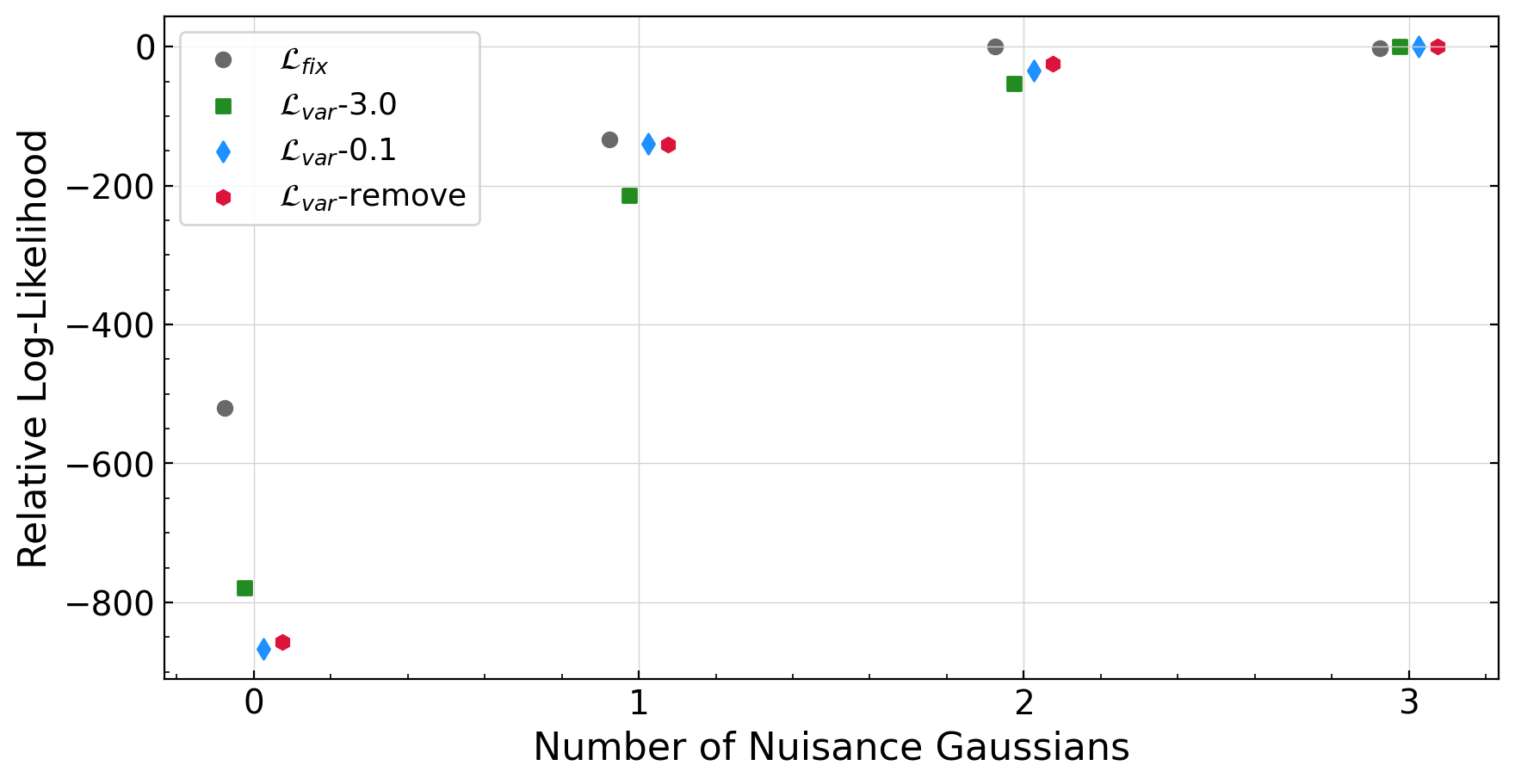}
    \caption{Log of the likelihood ratio between runs with $3$ nuisance Gaussians and those with fewer nuisance Gaussians, separately for each of the four likelihood choices.  The posteriors for each run are shown in in Fig.~\ref{fig:apr06hi_results} above.  This establishes that the fit improves as more nuisance parameters are added.}
    \label{fig:log-likelihoods}
\end{figure}

By this measure, our modeling is successful.  Fig.~\ref{fig:log-likelihoods} shows the relative log-likelihood as the number of nuisance parameters is increased.  That is, for each choice of likelihood function, we present the difference between the 3-Gaussian best-fit log-likelihood and the $n$-Gaussian best-fit log-likelihood, as $n$ increases from 0 to 3.  This is the log of the ``likelihood ratio,'' a classical test for whether adding additional parameters has improved the fit.  We see that the fit indeed improves with increasing nuisance parameters, with diminishing returns in moving from $2$ to $3$.  As the posteriors for our parameters of interest are also relatively stable at this point, we do not consider more than $3$ nuisance Gaussians.

\begin{figure}
    \centering
    \includegraphics[width=\linewidth]{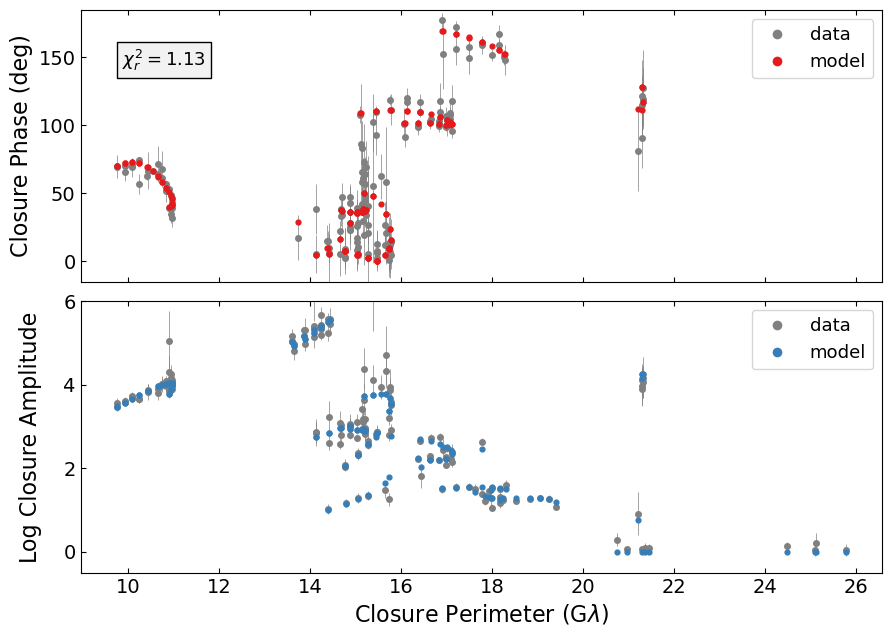}
    \caption{Visual agreement of model and data for the best fitting parameters of a fixed-likelihood run   (Fig.~\ref{fig:images} top).  We plot the absolute values of the closure quantities (the overall sign is arbitrary) as a function of the closure perimeter, defined to be the sum of the lengths of the baselines involved.  The error bars show (twice) the square root of the diagonal terms in the covariance matrices, i.e., $\sqrt{(\bm{\Sigma_\psi})_{nn}}$ for the $n^{\rm th}$ closure phase and $\sqrt{(\bm{\Sigma_c})_{mm}}$ for the $m^{\rm th}$ closure log-amplitude, computed using the data amplitudes $|\hat{V}_{ij}|$.  This fit has a reduced chi-squared value of $\chi^2_r = 1.13$, as compared with $1.06$ reported for the same dataset in \citetalias{EHT6}.  We emphasize that the visual agreement and reduced chi-squared value are \textit{not} indicative of the quality of fit in a statistical sense; rather, they simply give a rough idea of the degree of numerical agreement between data and best-fit model.}
    \label{fig:bestfit_fit}
\end{figure}

In its geometric modeling, EHTC reported a reduced chi-squared statistic defined by
\begin{align}
    \chi^2_{r} = \frac{1}{N_\psi + N_c - N_p} \left( \sum_{i=1}^{N_\psi} \frac{(\Delta \Psi_i)^2}{(\bm{\Sigma_\psi})_{ii}}+\sum_{j=1}^{N_c} \frac{(\Delta C_j)^2}{(\bm{\Sigma_c})_{jj}} \right), 
\end{align}
where $N_\psi$ is the number of closure phases, $N_c$ is the number of closure (log-)amplitudes, $N_p$ is the number of model parameters, and the variances $(\bm{\Sigma_\psi})_{ii}$ and $(\bm{\Sigma_c})_{jj}$ are computed in the ``fixed'' approximation (using measured amplitudes $|\hat{V}_{ij}|$ instead of mean values $|\bar{V}_{ij}|$ in Eq.~\eqref{variance_matrix}). In order to compare results, we have computed this statistic for our best-fit images, finding values broadly consistent with those reported by ETHC---an example is shown in Fig.~\ref{fig:bestfit_fit}.  However, we emphasize that the numerical value of this quantity does not indicate the absolute quality of the fit; we discuss it only as a ``code check'' comparison with EHTC results.\footnote{When the (non-reduced) chi-squared statistic is constructed from independent Gaussian random variables with known variances, it will be distributed according to the chi-squared distribution and can be converted to a $p$-value by comparing to the cumulative distribution function of the chi-squared distribution (i.e., from a ``$p$-value table'').  These conditions are not met for our model---see discussion at the start of this subsection.  Furthermore, even if the chi-squared had this kind of statistical interpretation, the idea that the reduced chi-squared should be greater than $1$ to avoid over-fitting would be inappropriate for this nonlinear model \citep{chi2}.}

\section{Just add one}\label{app:just_add_one}

Although the analytic solution of the null geodesic equation in the Kerr spacetime  is rather intricate \citep{gralla-lupsasca2020exact}, certain features can be understood analytically in simple terms.  For example, \citet{gralla-lupsasca2020lensing} noticed that, for an observer on the spin-axis of a Kerr black hole and for emission from the spin-equator, the arrival impact parameter is related to the emission radius by the simple instruction to ``just add one''.  
Here we generalize this observation to emission from a whole front/midplane region (colored green in Fig.~\ref{fig:conchy}) of an observer with arbitrary inclination relative to the spin axis.  Working in dimensions of black hole mass (i.e., all lengths are divided by $GM/c^2$), the formula is
\begin{align}\label{just_add_one}
    b \approx (r+1) \sin \theta',
\end{align}
where $(r,\theta,\phi)$ are (dimensionless) Boyer-Lindquist coordinates, while $\theta'$ is the polar angle relative to an observer at $(\theta_o,\phi_o)$, given by
\begin{align}
\cos \theta' = \cos \theta_s \cos \theta_o + \sin \theta_s \sin \theta_o \cos(\phi_o-\phi_s),
\end{align}
with $s$ for source and $o$ for observer.
In numerical experiments, we find that Eq.~\eqref{just_add_one} holds with $20\%$ accuracy or better for direct (non-orbiting) photons in the foreground/midplane region, for any spin and observer inclination.  More precisely, we generate random initial conditions for photons in the region $r<10$ surrounding the black hole and numerically solve the null geodesic equation.  For the subset of trajectories reaching infinity, we record 
the arrival angles $(\theta_o,\phi_o)$, the impact parameter $b$, and the total bending angle $\delta$.  The arrival angles are just the asymptotic (large-$r$) values of the Boyer-Lindquist coordinates, while the impact parameter is given as $b=\sqrt{\alpha^2+\beta^2}$ in terms of the screen coordinates $\alpha$ and $\beta$ of the observer (e.g., \citet{gralla-lupsasca2020lensing}).  We define the bending angle by treating the spatial trajectory of the null geodesic as a space curve in three-dimensional Euclidean space, defined by identifying the Boyer-Lindquist coordinates with spherical coordinates.  The total bending is then given by
\begin{align}
    \delta = \int \kappa ds,
\end{align}
where the integral extends along the curve from source to observer, with $\kappa$ the curvature and $s$ the arclength.  We find that Eq.~\eqref{just_add_one} holds with $20\%$ accuracy or better for $99\%$ of all trajectories with $\theta'<120\degree$ (foreground/midplane region) and $\delta<90\degree$ (non-orbiting).

We will continue to refer to the approximation \eqref{just_add_one} as ``just add one''.  It means that for sources in the foreground/midplane region of a distant observer, one can simply consider $r+1$ to be a spherical coordinate and use flat spacetime intuition.  Picking a source point in the green region of Fig.~\ref{fig:conchy}, we add 1 in the radial direction and then move in a straight line to the right to arrive at the approximate impact parameter of that photon.

In the special case $\theta_o=0$ and $\theta'=\pi/2$, the ``just add one'' formula holds with higher accuracy and can in fact be derived as an  analytic approximation at large radius  \citep{gates-hadar-lupsasca2020}.  We do not have any corresponding analytic derivation for the more general formula~\eqref{just_add_one}, which remains a purely numerical observation.


\section{From source to ring}\label{app:mechanisms}

In this appendix we a provide a semi-quantitative exploration of the relationship between source properties and observed ring width.  We imagine a disk-like source of emission that is slightly inclined from a face-on orientation, which provides an excellent approximation to the state of the art source models \citep{gralla-lupsasca-marrone2020, chael-johnson-lupsasca2021}.  Since we are concerned primarily with ring width and not brightness asymmetry, we will neglect Doppler effects due to the inclination, working with a disk that is precisely face-on relative to the observer. 

In this case the emission is occurs at $\theta'\approx \pi/2$, and from  Eq.~\eqref{just_add_one}, noting that $b=\alpha/\alpha_M$, the observed angular radius $\alpha$ is
\begin{align}\label{ray-tracing-is-hard}
    \alpha/\alpha_M \approx r + 1,
\end{align}
where $\alpha_M$ is defined in Eq.~\eqref{thetaM} above, and $r$ is the Boyer-Lindquist radius of emission divided by $GM/c^2$.  Incorporating redshift and photon ring effects based on intuition from toy emission profiles \citep{gralla-holz-wald2019, gralla-lupsasca-marrone2020}, we arrive at the following rules of thumb for identifying the emission region associated with observed radiation.  For radiation observed at angular radius $\alpha/\alpha_M \gtrsim 6$, Eq.~\eqref{ray-tracing-is-hard} can be used directly to infer a rough emission profile $I_{\rm em}(r)$ from an observed profile $I_{\rm obs}(\alpha)$.  For $4.5 \lesssim \alpha/\alpha_M \lesssim 5.5$, the radiation comes both from direct photons leaving from the corresponding radius \eqref{ray-tracing-is-hard} and also from orbiting photons arriving from all emission radii (the photon ring).  The observed radiation will be roughly twice as bright as one would naively infer from a smooth emission profile.  For $\alpha/\alpha_M \lesssim 4$, the observed radiation will be dimmer than one would naively infer,  due to strong gravitational redshift at $r \lesssim 3$.  The combination of these effects means that emission peaked near the horizon gives rise to a ring whose blurred brightness peaks near $\alpha/\alpha_M \approx 5$, just as if the emission had all originated at $r \approx 3$--$4$ (see, e.g., \citet{gralla-holz-wald2019}).  

If the gas-dynamics mass $\psi \approx 0.5$ is assumed, then the observed ring radius of $\alpha\approx 20\mu$as is $11 \alpha_M$, corresponding to emission from $r \approx 10$.  General-relativistic effects are negligible, and the width of the ring can be transcribed into an emission profile.  If the fractional width is reasonably large, such an emission profile could be associated with the retrograde innermost stable circular orbit (ISCO) of a high-spin black hole ($r=9$).  A narrow ring is by contrast highly implausible, as one would need a mechanism for a narrow emission profile near $r=10$. 

For a black hole of mass $\psi=0.75$ (intermediate between the gas-dynamics and stellar-dynamics values), the observed ring radius $\alpha\approx 20\mu$as would correspond to emission from $r\approx6$.  Again, one can easily imagine a broad emission region associated with the ISCO or some other radius, and indeed this type of profile is seen among EHTC models at all spin \citepalias[Fig.~2, top row]{EHT5}.\footnote{EHTC reported a mass measurement corresponding to $\psi=1.05 \pm 0.1$ based on its suite of source models.  The models that appear to be consistent with lower masses like $\psi=0.75$ \citepalias[Fig.~2, top row]{EHT5} were largely excluded from consideration based on a lack of Poynting flux in the associated GRMHD simulations \citepalias[Table 2]{EHT5}, as compared to the observed power in the M87 jet.}  However, a narrow ring from a black hole of mass $\psi=0.75$ remains implausible, as there is no mechanism (beyond an unnaturally narrow emission profile) to produce one. 

If the stellar-dynamics mass $\psi=1$ is assumed, then the observed ring radius $\alpha\approx 20\mu$as is approximately $5.5 \alpha_M$, and relativistic effects come in to play.  As described above, the combination of dimming from gravitational redshift and brightening from orbiting photons compresses a smoothly varying emission profile into a thinner structure observed around $5 \alpha_M$.  While this easily permits somewhat narrow fractional widths (say, $\sim30\%$), it remains difficult to imagine the range $10$--$20\%$ seen in Fig.~\ref{fig:rings}.  
For example, a source model recently analyzed by \cite{chael-johnson-lupsasca2021} has emission dropping off extremely rapidly from the horizon, such that it is reduced by a factor of 5 already at $r=3$ (Fig.~4 therein).  Even this already-narrow emission profile, coupled with the narrowing effects of strong-field lensing, produces a ring whose visually-estimated fractional width\footnote{The FWHM is not a suitable definition of width for models including the photon ring, since the photon ring brightness diverges logarithmically (cut off by optical depth).  In making these visual estimates we imagine blurring the image by at least the width of the photon ring.  It is not obvious that such an estimate can be compared to the FWHM of a ring model fit to EHT data.  However, some evidence in favor of this association is provided by Figs.~12 and 16 of \citet{psaltis2020}, showing synthetic-data examples where the the FWHM of the best-fit ring corresponds to the broader (non-photon ring) emission in the true underlying image.} (Figs.~1 and 7 of \citeauthor{chael-johnson-lupsasca2021}) is at least 25\%, and which looks quite dissimilar to the narrow rings shown in Fig.~\ref{fig:rings}.


In summary, generating such narrow rings would require a departure from the conventional picture of the origin of the 1.3mm emission. To emphasize this point, we now discuss---and largely dismiss---four theoretical ideas for producing very thin rings.

One possibility is that the dominant ``direct'' (non-lensed) component of emission could be outside the effective field of view, such that \textit{only} the narrow photon ring is seen.  However, an interferometer does not ``aim'' at the center of its image like an ordinary telescope; instead, the correlation process defines an image center associated with the brightest fringes.  This process would undoubtedly have picked up the dominant direct emission on most EHT baselines, and would not produce an image with only a photon ring.

A second effect capable in principle of producing a thin ring is an Einstein ring from a bright, localized source behind the black hole.  One could imagine that the $1.3$mm emission is dominated by a bright, orbiting ``hot spot'' which happened to pass behind the black hole with perfect alignment during the 2017 observational campaign.  However, this would seem to be inconsistent with previous measurements finding roughly consistent levels of horizon-scale 1.3mm flux density \citep{wielgus-etal2020}.

A third possibility is that the emission arises from the ergoregion of the black hole.  Although the spacetime metric varies smoothly, the physical properties relative to asymptotic infinity change dramatically in the ergoregion.  In the context of force-free or magnetohydrodynamic plasmas, the relevant feature is the introduction of an \textit{inner light surface}, inside of which plasma and energy can only flow inwards toward the black hole \citep{komissarov2004,gralla-jacobson2014}.  The existence of this surface also poses an obstacle to global continuity of magnetic field lines, causing current sheets to form near the spin-equator of the black hole \citep{komissarov2004,nathanail-contopoulos2014,parfrey-philippov-cerutti2019}.  These current sheets are natural candidates for bright emission, which would be observed as a thin ring.  However, this near-horizon emission would be strongly redshift-dimmed.

A fourth possibility is that some combination of global dynamics, local plasma effects, and/or observer orientation conspires to produce a narrow ring.  For example, GRMHD simulations that contain collimated Poynting-flux outflows tend to show a sharp boundary between a ``jet region'' where the density is very low, and a broader ``wind region'' where it more closely matches conditions in the disk (e.g., \citeauthor{nakamura2018} \citeyear{nakamura2018}).  This gives rise to a rather sharp change in plasma properties across a funnel-shaped ``sheath'' surrounding the base of the jet.  If some unknown process causes $1.3$mm emission to become concentrated in this area, the result would be a conical or paraboloidal sheet of emission.  If the emission is entirely in the sheath near the black hole, and if furthermore the jet opening angle is small and the observer is nearly aligned with the jet, then the observational appearance could be a thin ring.  One might also expect special emission features at the stagnation surface marking the boundary between inflow and outflow \citep{mckinney2006,broderick-tchekhovskoy2015}, where unscreened electric fields are expected to accelerate particles.  A very thin ring could be produced if the emission is concentrated at the \textit{intersection} of the stagnation surface and the jet base sheath.

None of these potential theoretical explanations for a very thin ring is particularly promising, as they generally require a conspiracy of different effects, an unnatural fine-tuning, and/or a surprising concentration of the emission in one region.


\bsp	
\label{lastpage}
\end{document}